\documentclass[review]{elsarticle}
\usepackage[margin=2.3cm]{geometry}
\usepackage{lineno}
\usepackage[dvipsnames]{xcolor}
\modulolinenumbers[5]
\usepackage{footmisc}
\usepackage{diagbox}
\makeatletter
\def\ps@pprintTitle{%
 \let\@oddhead\@empty
 \let\@evenhead\@empty
 \def\@oddfoot{}%
 \let\@evenfoot\@oddfoot}
\makeatother

\usepackage{lipsum}
\usepackage{bmpsize}
\usepackage{graphicx}
\usepackage{setspace}

\usepackage{todonotes}

\usepackage[utf8]{inputenc}
\usepackage[T1]{fontenc}
\usepackage{newtxtext,newtxmath}
\usepackage{siunitx}
\usepackage{pdfpages}
\usepackage{amsmath}
\usepackage{amsfonts}
\usepackage{pifont} 
\usepackage{bbm}
\usepackage{calligra} 
\usepackage{paralist}
\usepackage{subfigure}
\usepackage{float}
\usepackage{nomencl}

\usepackage[final]{changes} 

\makeatletter 
\renewcommand\subsection{\@startsection{subsection}{2}{\z@}%
           {12\p@ \@plus 6\p@ \@minus 3\p@}%
           {3\p@ \@plus 6\p@ \@minus 3\p@}%
           {\normalfont\normalsize\bfseries\itshape}}
\renewcommand\subsubsection{\@startsection{subsubsection}{3}{\z@}%
             {-3.25ex\@plus -1ex \@minus -.2ex}%
             {-1.5ex \@plus -.2ex}
             {\normalfont\normalsize\bfseries\emph}}
                                     
\renewcommand\paragraph{\@startsection{paragraph}{4}{\z@}%
            {-3.25ex\@plus -1ex \@minus -.2ex}%
             {-1.5ex \@plus -.2ex}
            {\normalfont\small\bfseries\texttt}}
\makeatother

\begin{document}

\begin{frontmatter}

\title{Occupant Plugload Management for Demand Response in Commercial Buildings: Field Experimentation and Statistical Characterization}

\makeatletter
\def\@author#1{\g@addto@macro\elsauthors{\normalsize%
    \def\baselinestretch{1}%
    \upshape\authorsep#1\unskip\textsuperscript{%
      \ifx\@fnmark\@empty\else\unskip\sep\@fnmark\let\sep=,\fi
      \ifx\@corref\@empty\else\unskip\sep\@corref\let\sep=,\fi
      }%
    \def\authorsep{\unskip,\space}%
    \global\let\@fnmark\@empty
    \global\let\@corref\@empty  
    \global\let\sep\@empty}%
    \@eadauthor={#1}
}
\makeatother



\author{Chaitanya Poolla\corref{correspauthor}\fnref{chaitanya_affiliation}}
\fntext[chaitanya_affiliation]{Chaitanya Poolla is with Intel Corporation. This work was done when he was with Carnegie Mellon University (SV).}

\author{Abraham K. Ishihara\fnref{abe_affiliation}}
\fntext[abe_affiliation]{Abraham K. Ishihara is with KBR. This work was done when he was with Carnegie Mellon University (SV).}

\author{Dan Liddell\fnref{dan_affiliation}}
\fntext[dan_affiliation]{Dan Liddell contributed to this work when he was with Carnegie Mellon University (SV).}

\author{Rodney Martin\fnref{rodney_affiliation}}
\fntext[rodney_affiliation]{Rodney Martin is with NASA Ames Research Center.}

\author{Steven Rosenberg\fnref{steve_affiliation}}
\fntext[steve_affiliation]{Steven Rosenberg is with Carnegie Mellon University (SV).}

\cortext[correspauthor]{Corresponding author: \emph{cpoolla@alumni.cmu.edu}}
\begin{abstract}
Commercial buildings account for approximately 35\% of total US electricity consumption, of which nearly two-thirds is met by fossil fuels resulting in an adverse impact on the environment. This adverse impact can be  mitigated by lowering energy consumption via control of occupant plugload usage in a closed-loop building environment. In this work, we conducted multiple experiments to analyze changes in occupant plugload energy consumption due to incentives and/or visual feedback. The incentives entailed daily monetary values between \$5 and \$50 administered in a randomized order and the visual feedback consisted of a web-based dashboard aimed at increasing the energy awareness of participants. Experiments were performed in government office and university buildings at NASA Ames Research Park located in Moffett Field, CA. Autoregressive models were constructed to predict expected plugload savings in the presence of exogenous variables. Analysis of the data revealed modulation of plugload energy consumption can be achieved via visual feedback and incentive mechanisms suggesting that occupant-in-the-loop control architectures may be effective in the commercial building environment. Our findings indicate that the mean energy reduction due to visual feedback in office and university environments were $\approx 9.52\%$ and $\approx 21.61\%$, respectively. By augmenting the visual feedback in the university environment with a monetary incentive, the mean energy reduction was found to be $\approx 24.22\%$.
\end{abstract}


\end{frontmatter}


\section{Introduction}
\label{intro}
Buildings account for more than $55\%$ of electricity consumption and $28\%$ of energy-related $CO_2$ emissions worldwide \cite{iea2019buildingsrole}. In 2019, these emissions amounted to $10$ Gigatonnes, reaching their highest level ever recorded \cite{iea2020buildingstracking}. Considering the rapid growth in building energy demand and related $CO_2$ emissions, there is a need to improve energy efficiency measures in buildings for achieving a Sustainable Development Scenario (SDS) in the society \cite{iea2020buildingstracking, hashempour2020energy, longo2019review}. Within the US, electricity consumed by residential, commercial, and industrial buildings account for $37\%$, $35\%$, and $28\%$, respectively \cite{useia2020annual}. 

Electricity consumption in buildings is classified into three categories: Heating, Ventilation, and Air Conditioning (HVAC); Lighting; and Plug and Process Loads (PPLs) \cite{nrelplugloads}. HVAC and lighting systems contribute to nearly two-thirds of the electricity consumption and PPLs account for the remainder \cite{mckenney2010commercial}. Building electricity consumption is reduced through methods such as Demand Response (DR) programs \cite{sehar2017integrated} or retrofitting \cite{castleton2010green}. While there have been several studies on reducing the energy consumption incurred by HVAC and lighting systems \cite{salsbury2005survey,wen2011personalized}, the problem of plugload energy reduction has received considerably less attention despite being recognized as the next big hurdle toward improving energy efficiency in buildings \cite{kaneda2010plug,agdas2015energy}.

The potential for energy savings due to plugload energy reduction in office buildings is estimated to be in the range of $15-40\%$ \cite{mercier2011commercial,acker2012office}. Given that plugloads are controlled by occupants, it is of interest to determine how interventions motivate occupants to vary their plugload energy consumption. Previous studies indicate that promoting energy efficiency among occupants by feedback and/incentives results in significant energy savings: Jain et al. \cite{jain2012assessing} studied the role of interventions in motivating energy efficient behavior among 43 participants over a period of six weeks. They concluded that feedback via historical comparisons and incentives are statistically significant in motivating energy reduction. Based on techniques from human-computer interaction, psychology, and energy efficiency, Yun et al. \cite{yun2013sustainability} found that $12\%-20\%$ energy savings can occur by behavioral modification. Petersen et al. \cite{petersen2007dormitory} conducted a five-week study involving feedback and incentives across 18 dormitories and found a significant energy conservation of $32\%$. The potential of online games for plugload energy conservation was studied by Gandhi et al. \cite{GANDHI20161}. Online games were found effective to reinforce conservation behavior, while suggesting the need for further studies on non-monetary commercial plugload reduction. \\ 
Despite its significance toward building energy efficiency, the problem of plugload management is challenging due to its dependence on occupant usage \cite{kaneda2010plug,agdas2015energy}. This dependence also results in the difficulty to model plugload consumption. Nevertheless, models are essential to forecast energy consumption and to assess the efficacy of interventions \cite{bourdeau2019modeling}. In most of the published literature on plugload energy consumption, either field experimentation or statistical modeling are absent. This absence limits the scope for demand management and sustainable development in cities. Thus, there exists a need for plugload studies that not only involve experiments but also infer models from the resulting data. To the best of our knowledge, this is the first work to statistically characterize and model occupant plugload energy consumption in commercial buildings based on designed experiments involving incentives and/or visual feedback.

In this work, two field experiments were designed to study the effect of incentives and/or dashboard feedback on occupant plugload energy consumption. These experiments were carried out in government office and educational (university) buildings inside the NASA Ames Research Park Complex. The experiment related to the government office building was conducted within NASA Sustainability Base (SB), a 50,000 square foot LEED platinum certified commercial building \cite{nasa_sb} in the presence of a visual (dashboard) feedback intervention. The other experiment was conducted within two Carnegie Mellon University Silicon Valley (CMU SV) campus buildings (NASA Ames Research Park buildings 19 and 23) in the presence of incentives and/or dashboard feedback. The data from these experiments was used to construct models for predicting the effects of incentive and feedback interventions on plugload energy consumption. These occupant plugload models enable the integration of occupants-in-the-loop within existing building optimization frameworks. The integration of the occupant plugload allows for holistic load modeling, thereby enabling optimal energy management in buildings. For example, a model-based optimal policy could be designed and updated in real-time based on system identified parameters. \cite{poolla2019designing}. The major contributions of this work are:
\begin{enumerate}
\item A paired experiment design to study the effects of dashboard-enabled feedback and/or incentives on occupant plugload energy consumption.
\item A statistical characterization of the data where hypothesis tests are conducted and confidence intervals are estimated to determine the efficacy of interventions. Autoregressive models with exogenous inputs are proposed to model occupant plugload energy consumption. This can enable occupants-in-the-loop control strategies within a demand response framework.
\item Visual feedback via informative dashboards offers statistically significant plugload reduction in both government office and university environments.
\end{enumerate}
The rest of this paper is organized as follows: Section \ref{plugload_experiment} describes the design and execution of the experiments at NASA SB and CMU SV. A statistical analysis of the data is presented in Section \ref{plugload_stat_model} along with the respective results and discussion. Concluding remarks are presented in Section \ref{conclusion}.

\section{Experiment design and execution}
\label{plugload_experiment}
We designed and conducted experiments to study the influence of incentives and/or feedback interventions on occupant plugload energy consumption. Our research hypothesis is:
\begin{itemize}
\item[]\emph{Providing incentives and/or dashboard-based feedback to occupants in \deleted{commercial buildings}\added{office and university buildings} reduces occupant plugload energy consumption.}
\end{itemize}
Consequently, we examine the claim that \emph{the average occupant plugload energy consumption in the presence of an incentive and/or feedback is less than the energy consumption in the absence of incentive and/or feedback} based on data from the experiments. In the rest of this section, we present the experiment design and implementation.
\subsection{Location and duration}
\label{expt_location}
Two experiments were conducted within the NASA Ames Research Center, one within a government office environment (NASA SB) and the other within a university environment (CMU SV - buildings 19 and 23). Let the symbols $\mathbb{E_N}$ and $\mathbb{E_C}$ denote the experiments at NASA SB and CMU SV, respectively. Each experiment consisted of multiple phases classified into a baseline or an experiment phase depending on the absence or presence of an intervention, respectively. The interventions employed during experiment phases consisted of \emph{dashboard feedback}, \emph{incentives}, or \emph{both dashboard feedback and incentives}. 
\added{The feedback-based intervention was applicable to both experiments $\mathbb{E_N}$ and $\mathbb{E_C}$ whereas the incentive-based intervention was applicable only to experiment $\mathbb{E_C}$ (at CMU SV). Thus, experiment $\mathbb{E_C}$ was designed to study both the individual and interaction effects of the interventions using a two-factor factorial design. The duration of both of the experiments were based on findings from habit modeling \cite{lally2010habits}}. The properties for each phase of $\mathbb{E_N}$ and $\mathbb{E_C}$ are provided in table \ref{table:expt_interventions}.
\begin{table}[H]
\centering
\resizebox{0.75\columnwidth}{!}{%
    \begin{tabular}{| l | c | c | c | c |}
    \hline
    \diagbox{\textbf{Expt. Property}}{\\[-0.5ex] \textbf{Phase}} & Baseline Phase & Incentive Phase & Feedback Phase & Feedback \& Incentive Phase \\ \hline
    $\mathbb{E_N}$ (NASA SB): Applicable? & Yes & No & Yes & No \\ \hline
    $\mathbb{E_N}$ (NASA SB): Notation& $\mathcal{P}1^\mathbb{N}$ & N/A & $\mathcal{P}3^\mathbb{N}$ & N/A \\ \hline
    $\mathbb{E_N}$ (NASA SB): Duration & Five weeks & N/A & Four weeks & N/A \\ \hline
    $\mathbb{E_C}$ (CMU SV): Applicable? & Yes & Yes & Yes & Yes \\ \hline
    $\mathbb{E_C}$ (CMU SV): Notation & $\mathcal{P}1^\mathbb{C}$ & $\mathcal{P}2^\mathbb{C}$ & $\mathcal{P}3^\mathbb{C}$ & $\mathcal{P}4^\mathbb{C}$ \\ \hline
    $\mathbb{E_C}$ (CMU SV): Duration & Five weeks & Two weeks & Two weeks & Two weeks \\ \hline
    \end{tabular}%
}
    \caption{Description of properties in experiments $\mathbb{E_N}$ and $\mathbb{E_C}$}
    \label{table:expt_interventions}
\end{table}
\nomenclature{$\mathbb{N}$}{The plugload experiment at NASA SB}
\nomenclature{$\mathbb{C}$}{The plugload experiment at CMU SV}


\subsection{Variables}
\label{expt_design}
We discuss the response variables and interventions in both experiments here. The response variable was defined as the time-averaged power consumption of the participant. 
Its value was computed based on data from smart powerstrips \cite{enmetric}. The interventions employed are described in Table \ref{table:expt_interventions}. The incentive interventions were administered as daily monetary rewards aimed at promoting energy conservation among the participants. The feedback intervention was administered by a web browser-based dashboard tool which was designed to raise awareness about the participant's plugload energy consumption. It is important to note that the feedback provided by the experimenters and the feedback received by the participants differed since each participant does not necessarily use all features of the dashboard at all times. The time spent by each participant on their dashboard is used to quantify the feedback received.
\subsection{Design principles and implementation}
\label{expt_principles}
The ideal experiment design allows us to attribute any changes in energy consumption exclusively to the interventions employed and not to nuisance factors such as differences in participants' workloads or preferences. For example, one participant may consume more energy than another due to the difference in a device workload or setting. In such a case, the device workload and setting are the inevitable nuisance factors that the experiment design needs to account for. By accounting for nuisance factors, we can ensure a strong connection between the interventions employed and the response recorded. To this end, we employ a combination of the recommended design techniques consisting of blocking, randomization, and replication \cite{montgomery2008design} as described below:
\begin{enumerate}
\item The experiments at the office building ($\mathbb{E_N}$) and the university ($\mathbb{E_C}$) were regarded as separate experiments instead of a single experiment with larger sample size. This separation mitigates the influence of nuisance factors such as those arising from heterogeneous work environments or device setups.
\item The participants in each experiment were randomly selected without preferences toward age, gender, work function, or any other nuisance factors except for their willingness to participate\footnote{The sample is selected at random with respect to the population of potential participants than that of occupants. However, distinguishing these populations is beyond the scope of this study as it required voluntary consent for participation.}. Such randomization is necessary to average out the effect of such nuisance factors that cannot be blocked.
\item The participants in each experiment phase were paired to their baseline selves as the control counterparts. Such a matched paired design blocks the effect of inter-participant variation due to difference in workloads and/or energy consumption preferences.
\end{enumerate}
In addition to blocking and randomization, the sample size was so determined to allow for the study of statistical differences in the experimental responses. In this manner, the experiment was designed to ensure well-treated, random samples for the statistical analyses described in Section \ref{plugload_stat_model}.

\subsection{Feedback intervention design: Dashboard application}
\begin{figure}[h]
    \centering
    \includegraphics[width=\textwidth]{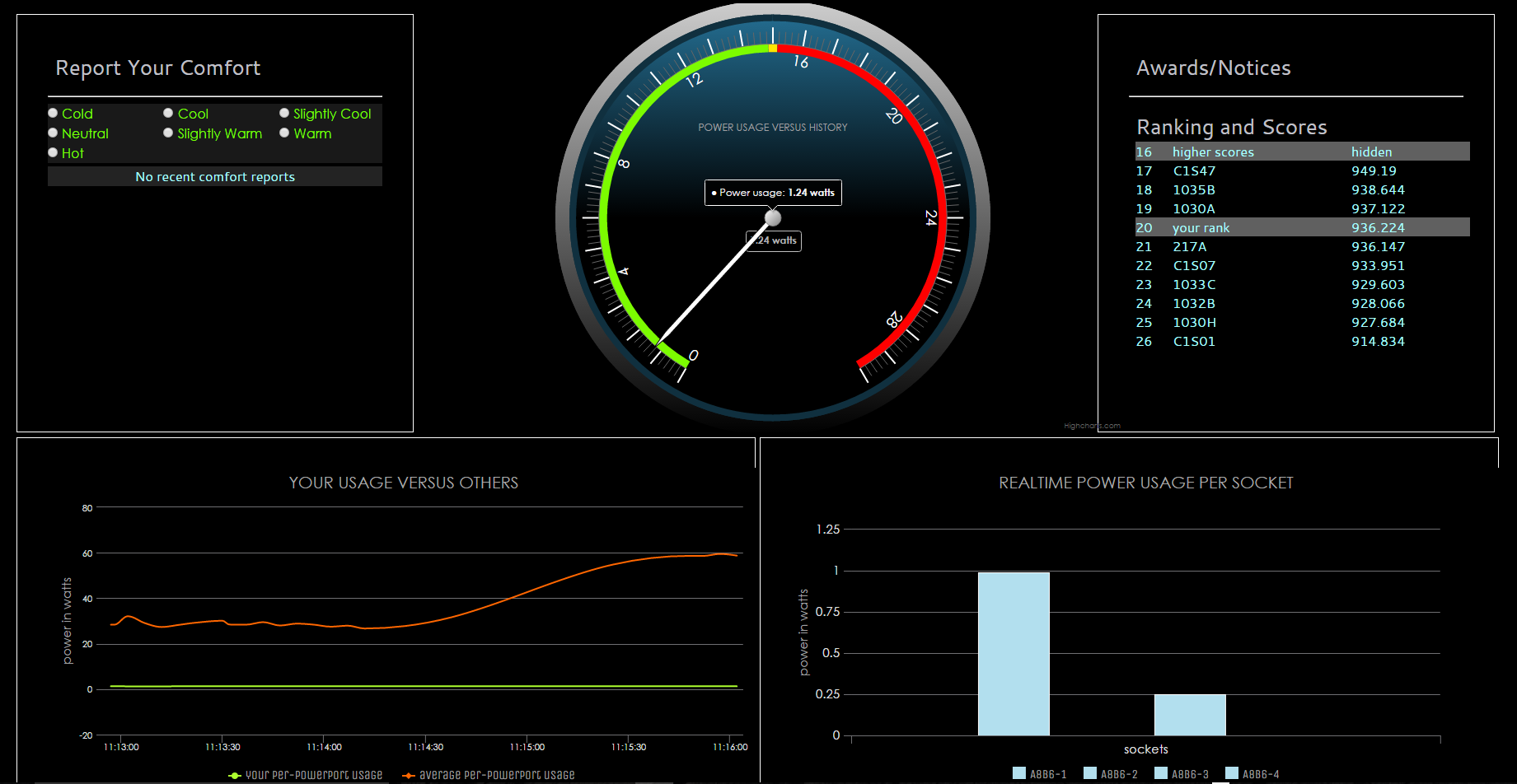}
    \caption{A screenshot depicting feedback provided via the dashboard}
    \label{FIG:dashboard}
\end{figure}

A dashboard was designed to provide the participants with information relevant to their plugload energy consumption. The elements of the dashboard were defined based on analytics that were previously found effective in motivating energy conservation among occupants in commercial buildings \cite{yun2013sustainability,yun2013toward,gulbinas2014effects}. These analytics were represented by easily comprehensible elements with minimal cognitive and visual load \cite{brath2004dashboard}. The back end of the dashboard was implemented in PHP and the front end was implemented as a browser-independent webpage in HTML, JavaScript, jQuery, and HighCharts. The servers for the webpage were hosted on Amazon AWS EC2 and RDS instances. An image of the dashboard is shown in Figure \ref{FIG:dashboard}.
Each feature of the dashboard is described in section \ref{sssec:graphical_elements}.
\subsubsection{Dashboard features}
\label{sssec:graphical_elements}
\begin{enumerate}
\item Comfort feature (upper left): The \emph{comfort feature} was represented by mutually exclusive radio buttons that allow participants to report their comfort levels. The options represented an ASHRAE 7-point scale \cite{de2002thermal}. This feature motivated participant engagement based on their historical interest in communicating their comfort levels to building facilities. Along with this feature, the dashboard also displayed energy consumption-related features.
\item Individual power feature (center): The \emph{instantaneous power consumed by the individual participant} was pointed to by the needle in the dial. Similar visualizations were found effective for energy reduction in households \cite{petkov2011motivating}. The dial's needle was set to saturate beyond the dial's maximum reading. The dial was calibrated using data collected during the baseline phase. The average baseline power usage was computed by considering data points above 5 W. This average was chosen to represent the zenith of the dial for the participant under consideration. The 5 W threshold was chosen to avoid a participant's inactivity from lowering the average value. The calibration also provided the color-coded context within which the current usage was positioned. 
\item Scoreboard feature (upper right): The \emph{scoreboard feature} provided participants with the score and relative position in the participant pool. When an incentive was provided, the participant with the highest score (rank 1) was declared the winner of the day. The scoring mechanism was designed to measure the improvement of the participant compared to his/her baseline, and is described in Section \ref{sssec:score_comp}.
\item Serial power feature (lower left): The power series of an individual (in orange) relative to the pool (in green) was depicted by line charts in the \emph{serial power feature}. Such social comparisons have proven successful in motivating energy reduction among participants \cite{allcott2011social,ayres2012evidence}. The vertical axis depicting power usage was scaled based on the individual and pool values during the time the dashboard window was active in the corresponding session.
\item Socket split feature (lower right): The instantaneous power consumed via the individual sockets in the powerstrip was represented by bar charts. While other features represented the participant's cumulative power consumption across sockets, these bars provided actionable feedback by corresponding to the device plugged in the socket.
\item Notification feature (top right): A \emph{notification feature} was provided in the dashboard to notify winners, if applicable.
\end{enumerate}
\subsubsection{Score computation}~\\
\label{sssec:score_comp}
\indent The scoreboard described above represents the participant's score along with the relative position in the competition against other participants to win the incentive. The steps involved in the scoring mechanism are described below:
\begin{enumerate}
\item The time-averaged power consumption across each powerstrip socket was computed for the baseline phase by excluding data points below an inactivity threshold (5 W). The threshold served as a measure of inactivity.
\item The socket-specific averages computed above were aggregated over all the sockets assigned to a participant to obtain the average active baseline power consumption of a participant.
\item The above steps were repeated across all participants to obtain baselines for the score computation described below.
\item During each day of the incentive competition, each participant's average active power consumption was determined similar to determining the baseline. The only procedural difference between the experiment and baseline computations was that the average power during the experiment was computed using data from local midnight till the scoring instant unlike the baseline computation which was performed using data from midnight to the next midnight.
\item The participant score was computed by the percentage improvement during the experiment compared to his/her baseline. That is, $score = 900 + 100\times\frac{baseline\_average-expt\_average}{baseline\_average}$, where $baseline\_average$ and $expt\_average$ represent the participant's baseline average power (step 2) and the experiment day average power (step 4), respectively. The choice of the score range in the interval [900, 1000] was motivated by previous findings that suggest a relationship between round numbers and human goals \cite{pope2011round}.
\end{enumerate}
\added{In this manner, the scoring mechanism assigns scores to the participants based on power reduction relative to their baseline consumption. We also note that an already energy-conscious participant may receive a low score due to a low baseline consumption and lack of ability for further energy reduction. This is expected since the study is designed to examine energy reduction irrespective of the baseline usage.}
\subsubsection{Inactivity detection}~\\
\label{sssec:inactivity_threshold}
\indent The inactivity threshold\footnote{The inactivity threshold was employed both in the scoring algorithm and the testing, estimation across all phases, thereby deeming any inferred power reduction conservative. \label{fnote:inactivity}} (5W) mentioned above was unknown to the participants to ensure that no participant was declared to be the winner either due to inactivity or absence. However, the participants were informed that the scoring mechanism only rewards reducing power consumption via active changes as opposed to reducing the power consumption via passive changes such as turning off devices or being absent. While turning off unused devices could be an active change, such a consideration could also encourage unwanted effects such as leaving devices off or working from other locations during the experiment just for the sake of incentives. Thus, we consider powering off devices to be passive and detect it using the inactivity threshold. Despite such inactivity measures, it was also possible that a participant could win due to apparent activity such as leaving a single unused device on while turning off all other devices. In such cases, a metric based on sliding time windows was used to detect participant inactivity. In this manner, the scoring algorithm was designed to guard against winning strategies driven by inactivity. 
\subsection{Incentive intervention design}
\label{inct_design}
For the experiment phases involving incentives ($\mathcal{P}2^\mathbb{C}$ and $\mathcal{P}4^\mathbb{C}$), a fixed monetary value was announced at the begin of each workday for participants to compete by changing their energy behavior compared to respective baselines. The values of the incentives ranged between \$5 and \$50 in multiples of 5 over a duration of ten working days or two weeks. Based on a random number generator, these values were randomly sampled without replacement to determine the incentive value for the day. The random ordering of incentives ensured that any changes in the consumption could not be attributed to the order in which the incentives were provided.


\subsection{Data collection}
\label{expt_data}
The power consumption of devices associated with each participant were monitored in real-time by smart powerstrips from Enmetric systems \cite{enmetric}. The monetary value associated with incentive inputs were recorded on a daily basis. As noted in \ref{expt_design}, the amount of feedback received by each participant was quantified by their time spent on the dashboard. In what follows, we refer to this time spent on the dashboard as the screentime. The screentime was measured as the aggregate duration spanned by the active sessions in the participant's browser. This duration was obtained by software running alongside the dashboard application.
\subsection{Execution of the experiment}
\label{expt_execution}
We describe the setup and implementation details for executing the experiment below.
\subsubsection{Experiment setup}~\\
\label{expt_setup}
\indent With the proposed design and permissions for experiments $\mathbb{E_N}$ and $\mathbb{E_C}$, the participants were recruited. At NASA SB, sixteen full-time employees were recruited for the experiment $\mathbb{E_N}$. At CMU SV (buildings 19 and 23), a mix of faculty, staff, and students totalling sixteen in number were recruited for experiment $\mathbb{E_C}$. Smart powerstrip(s) were installed in each participant's workspace for collecting data during baseline and experiment phases. The devices plugged into the powerstrips were noted as shown in Table \ref{table:device_list}.
\subsubsection{Experiment $\mathbb{E_N}:$ NASA Sustainability Base}~\\
\label{expt1_nasa_sb}
\indent This experiment was conducted in two phases, a baseline phase and a feedback intervention phase. The baseline phase ($\mathcal{P}1^\mathbb{N}$) spanned five weeks from September 12, 2016 to 17 October, 2016, during which no interventions were administered. Thereafter, the feedback intervention phase ($\mathcal{P}3^\mathbb{N}$) was conducted during which the participants were provided with dashboard feedback described in Section \ref{sssec:graphical_elements}. The participants were provided with relevant explanation as shown in Figure \ref{FIG:complete_dashboard_explanation}. This phase was conducted for four weeks from October 18, 2016 to November 11, 2016.
\nomenclature{$\mathcal{P}1^\mathbb{N}$}{Baseline phase of the NASA experiment}
\nomenclature{$\mathcal{P}3^\mathbb{N}$}{Feedback phase of the NASA experiment}
\subsubsection{Experiment $\mathbb{E_C}:$ CMU SV - buildings 19 and 23}
\begin{figure}[ht!]
     \begin{center}
        \subfigure[Display description provided during phase $\mathcal{P}2^\mathbb{C}$]{
        \centering
            \includegraphics[width=0.4\textwidth]{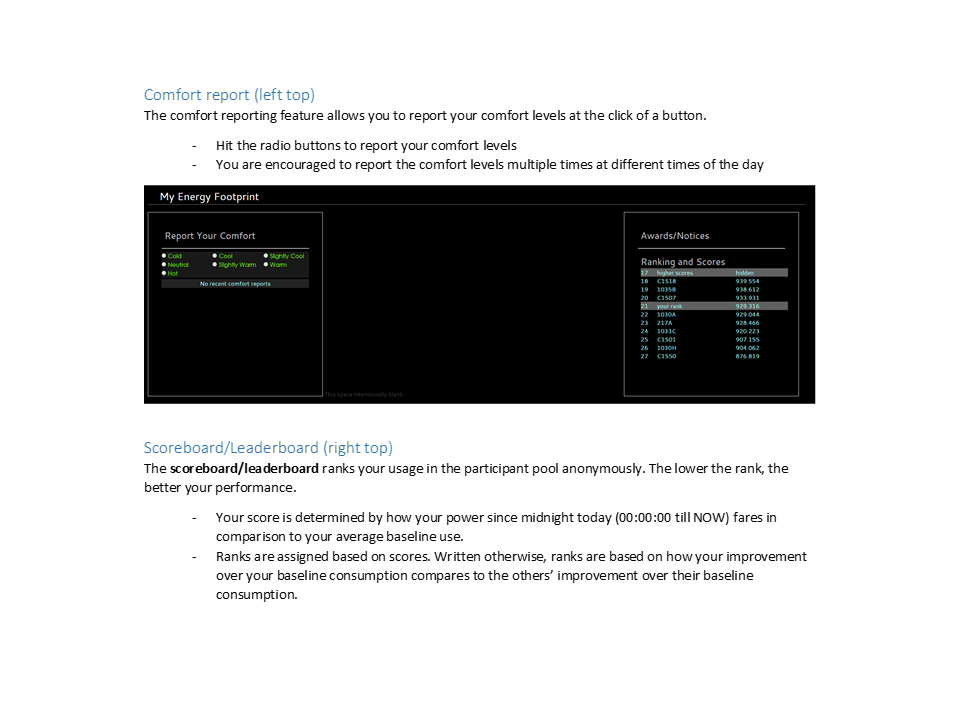}
            \label{FIG:incentive_only_dashboard_explanation}}\hfill
        \subfigure[Display description provided during phases $\mathcal{P}3^\mathbb{N}$ and $\mathcal{P}3^\mathbb{C}$]{
           \centering
           \includegraphics[width=0.4\textwidth]{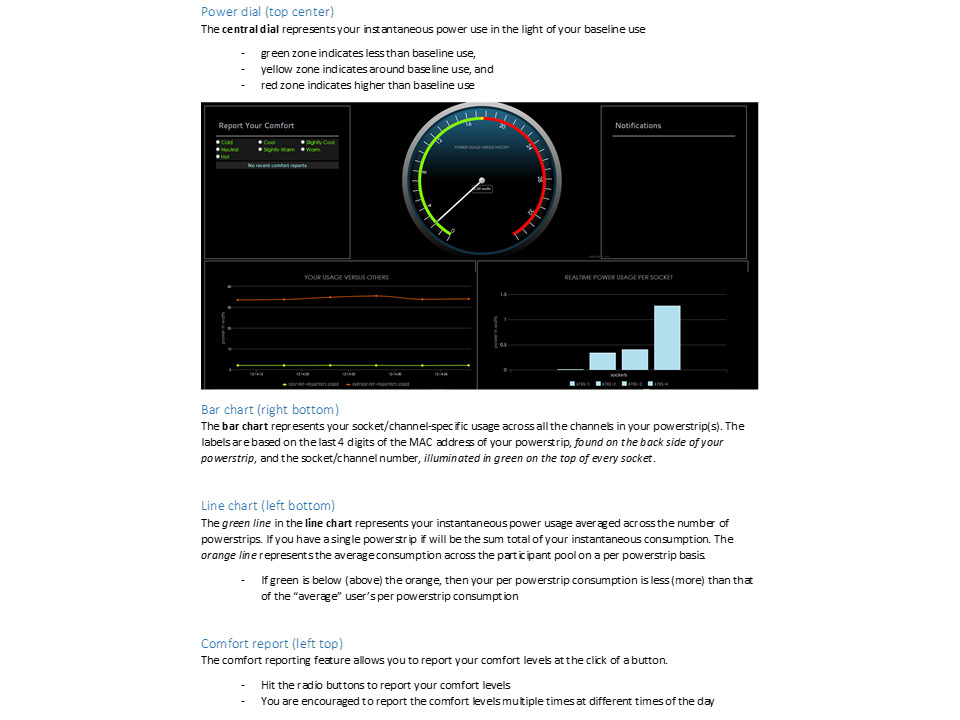}
           \label{FIG:feedback_only_dashboard_explanation}}
        \\ 
        \subfigure[Display description provided during phase $\mathcal{P}4^\mathbb{C}$]{
            \centering
            \includegraphics[width=0.4\textwidth]{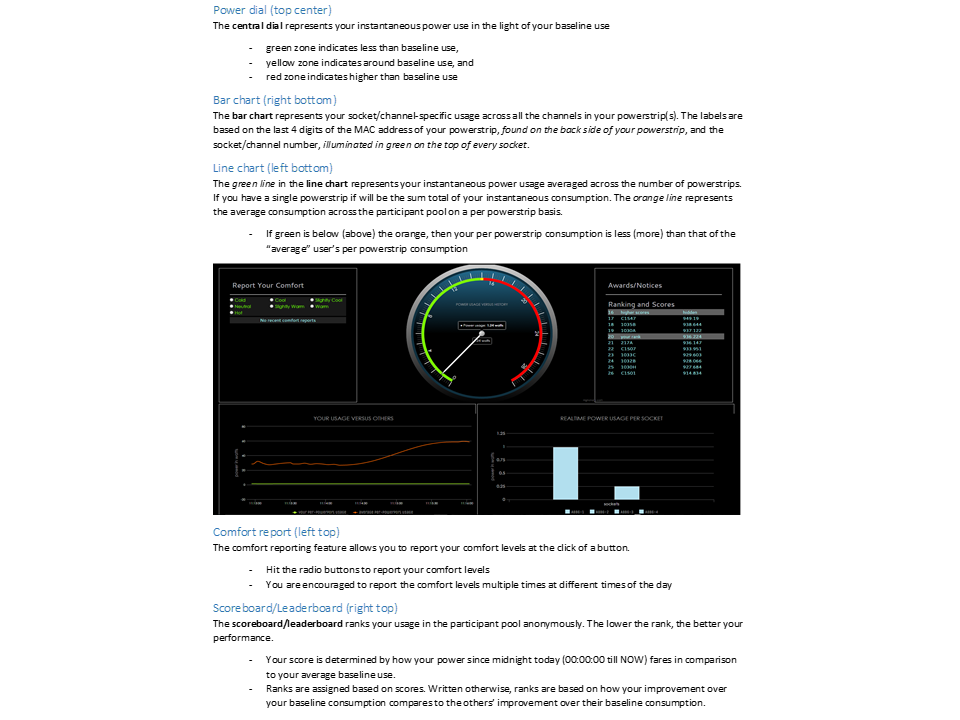}
            \label{FIG:complete_dashboard_explanation}}\hfill
        \subfigure[Information about possible energy conservation practices provided to participants during all the experiment phases]{
            \centering
            \raisebox{0.5\height}{\includegraphics[width=0.4\textwidth]{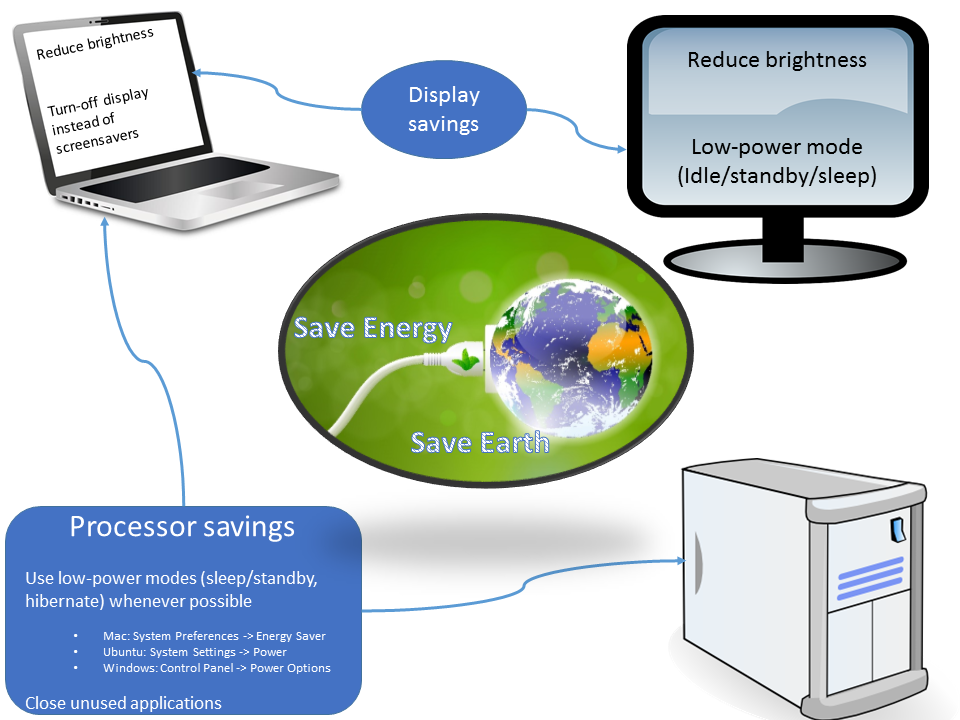}}
            \label{FIG:energy_conservation_practices}}\\
    \end{center}
    \caption{Information provided to the participants during the listed experiment phase(s)}
   \label{FIG:participantinfo}
\end{figure}

~\\
\label{expt2_cmu}
\indent The experiment $\mathbb{E_C}$ was conducted in four phases. The first phase was the baseline phase ($\mathcal{P}1^\mathbb{C}$) during which no intervention was administered. This phase was conducted for five weeks from September 12, 2016 to October 17, 2016. The second phase of the experiment was the \emph{incentive-only phase} ($\mathcal{P}2^\mathbb{C}$) wherein monetary incentives were provided for participants to compete with the objective of winning the incentive. The individual with the highest score at local midnight was the winner of the day considered. The participants were also provided access to dashboards containing only the scoreboard, which showed their near real-time scores. An explanation of the relevant elements received by the participants during this phase is shown in Figure \ref{FIG:incentive_only_dashboard_explanation}. The \emph{incentive-only} phase was conducted for two weeks from October 18, 2016 to October 30, 2016.
The third phase was the \emph{dashboard feedback only} phase ($\mathcal{P}3^\mathbb{C}$) during which each participant was provided with a dashboard depicting comparisons relative to their corresponding baseline and to the participant pool. All the dashboard features described in Section \ref{sssec:graphical_elements} except the scoreboard were provided to the participants. An explanation of the features shown in Figure \ref{FIG:feedback_only_dashboard_explanation} were provided to the participants. This phase was conducted for two weeks from October 31, 2016 to November 13, 2016. Finally, the \emph{both incentive and dashboard feedback} phase was conducted during which the participants were provided with both the incentive and dashboard feedback. All the features of the dashboard were made available to the participants during this phase. The participants were provided with explanations of each feature as shown in Figure \ref{FIG:complete_dashboard_explanation}. This phase was conducted for two weeks from November 14, 2016 to November 25, 2016.
\nomenclature{$\mathcal{P}1^\mathbb{C}$}{Baseline phase of the CMU experiment}
\nomenclature{$\mathcal{P}2^\mathbb{C}$}{Incentive-only phase of the CMU experiment}
\nomenclature{$\mathcal{P}3^\mathbb{C}$}{Feedback-only phase of the CMU experiment}
\nomenclature{$\mathcal{P}4^\mathbb{C}$}{Incentive+Feedback phase of the CMU experiment}
\subsubsection{Energy conservation information}~\\
\indent At the beginning of every experiment phase, namely $\mathcal{P}3^\mathbb{N}$, $\mathcal{P}2^\mathbb{C}$, $\mathcal{P}3^\mathbb{C}$, $\mathcal{P}4^\mathbb{C}$, the participants were provided with information on possible practices to reduce plugload energy consumption as shown in Figure \ref{FIG:energy_conservation_practices}. The specific choices on the appropriate implementations of these practices and compliance to safety regulations were left to the discretion of the participants. In this manner, the experimenters ensured that any absence of behavioral changes during the intervention phases could not be attributed to lack of information. These suggestions were compiled after surveying and classifying the devices used by each participant. The list of devices associated with experiments $\mathbb{E_N}$ and $\mathbb{E_C}$ are shown in Table \ref{table:device_list}.
\begin{table}
\centering
    \begin{tabular}{ | l | l | l | l | l | l | l | }
    \hline
    \emph{Experiment} & \emph{\#Monitors} & \emph{\#Laptops} & \emph{\#Docking stations} & \emph{\#Desktops} & \emph{\#Headsets} & \emph{\#Landline telephones}\\ \hline
    $\mathbb{E_N}$ & 21 & 11 & 9 & 2 & 7 & 9 \\ \hline
    $\mathbb{E_C}$ & 23 & 13 & 0 & 8 & 0 & 6 \\
    \hline
    \end{tabular}
    \caption{List of devices associated with experiments in NASA SB ($\mathbb{E_N}$) and CMU SV ($\mathbb{E_C}$)}
    \label{table:device_list}
\end{table}


\section{Statistical analysis and modeling}
\label{plugload_stat_model}
In this section, we analyze the results from experiments $\mathbb{E_N}$ and $\mathbb{E_C}$.
This analysis involves performing hypothesis tests, estimating confidence intervals, and developing statistical models from the data.
Given that each of the baseline and experiment phases for $\mathbb{E_N}$ and $\mathbb{E_C}$ were conducted over several days, we consider the following temporal context for analysis. Let the duration of each phase be represented as follows: For experiment $\mathbb{E_N}$, let the set of days corresponding to $\mathcal{P}j^\mathbb{N}$ be denoted by $D_j^\mathbb{N}$, where $j \in \{1,3\}$. Similarly, let the set of days corresponding to $\mathcal{P}k^\mathbb{C}$ be denoted by $D_k^\mathbb{C}$, where $k \in \{1,2,3,4\}$. In addition, let $\mathbb{D}_W = \{Monday, Tuesday, Wednesday, Thursday, Friday, Saturday, Sunday\}$ represent the days of a week.
Further, in a phase $\mathcal{P}j^X$ where $X \in \{\mathbb{N},\mathbb{C}\}$, let the set of participants be represented by $\mathcal{P}j^X_{All}$. Let the $i^{th}$ participant of this set be denoted by $\mathcal{P}j^X_i$.


\subsection{Data analysis for experiment $\mathbb{E_N}$ at NASA SB}
\label{ssec:nasa_analysis}
During the baseline phase $\mathcal{P}1^\mathbb{N}$, let the power consumption of the $i^{th}$ participant $\mathcal{P}3^\mathbb{N}_i$ on day $d \in D_1^\mathbb{N}$ at time instant $t$ be denoted by $y_{\mathcal{P}1^\mathbb{N}_i}(d,t)$ and let the time-averaged power consumption during $[t_0,t_f)$ be denoted by $\bar{y}_{\mathcal{P}1^\mathbb{N}_i}(d,t_0,t_f)$. Similarly, let the instantaneous and time-averaged power consumption during the feedback intervention phase $\mathcal{P}3^\mathbb{N}$ be denoted by $y_{\mathcal{P}3^\mathbb{N}_i}(d,t)$ and $\bar{y}_{\mathcal{P}3^\mathbb{N}_i}(d,t_0,t_f)$, respectively. Further, let this participant's screentime during the time interval $[t_0,t_f)$ on day $d$ be denoted by $x^A_{\mathcal{P}3^\mathbb{N}_i}(d,t_0,t_f)$. Let the random variable corresponding to the baseline response of the $i^{th}$ participant during the time interval $[t_0,t_f)$ on day $d \in D_1^\mathbb{N}$ be denoted by the random variable $Y_{\mathcal{P}1^\mathbb{N}_i}(d,t_0,t_f)$. Similarly, let the random variables associated with the response and screentime input during the feedback phase be represented by $Y_{\mathcal{P}3^\mathbb{N}_i}(d,t_0,t_f)$ and $X^A_{\mathcal{P}3^\mathbb{N}_i}(d,t_0,t_f)$, respectively.
\subsubsection{Statistical assumptions}~\\
\label{ssec:nasa_stat_assumptions}
\indent Given the baseline and experiment conditions, we represent the day of the week ($d \in \mathbb{D}_W$) statistics corresponding to $Y_{\mathcal{P}1^\mathbb{N}_i}(d,t_0,t_f)$ and $Y_{\mathcal{P}3^\mathbb{N}_i}(d,t_0,t_f)$ by $\mu_{\mathcal{P}1^\mathbb{N}_i}(d,t_0,t_f)$ and $\mu_{\mathcal{P}3^\mathbb{N}_i}(d,t_0,t_f)$, respectively. For hypothesis testing and interval estimation, we consider the sample constituted by the differences in the daily-averaged experiment response and the corresponding baseline response, sampled across participants and days of the week $\mathbb{D}_W$. The sample elements are assumed to be independent across the days of the week and the participant pool. Let this response differential for the $i^{th}$ participant on a day $d \in \mathbb{D}_W$ be represented by the random sample $\mu_{\mathcal{P}1^\mathbb{N}_i}(d,0,86400)-\mu_{\mathcal{P}3^\mathbb{N}_i}(d,0,86400)$. The response differential across participants and days of the week mitigates the corresponding nuisance factors stated in Section \ref{expt_principles}. Given these matched pairs, statistical testing allows us to attribute any significant changes between the baseline and experiment responses to the intervention administered rather than to nuisance factors such as differences in individual energy needs or differences due to daily work schedules.
\subsubsection{Hypothesis testing and confidence interval estimation}~\\
\label{ssec:nasa_hyp_test}
\indent We employ a paired difference test to examine the differential population sampled across participants and days of the week. Given the matched pairs $\mu_{\mathcal{P}1^\mathbb{N}_i}(d,0,86400)$ and $\mu_{\mathcal{P}3^\mathbb{N}_i}(d,0,86400)$, the paired difference t-test checks if the mean differential sample is significantly different from zero.
The null and the alternative hypotheses are presented below:
\begin{enumerate}
\item $\mathbf{H}_0^\mathbb{N}$: $\mu_{\mathcal{P}1^\mathbb{N}_i}(d,0,86400) - \mu_{\mathcal{P}3^\mathbb{N}_i}(d,0,86400)$ is sampled from a population with zero mean
\item $\mathbf{H}_A^\mathbb{N}$: $\mu_{\mathcal{P}1^\mathbb{N}_i}(d,0,86400) - \mu_{\mathcal{P}3^\mathbb{N}_i}(d,0,86400)$ is sampled from a population with non-zero mean
\end{enumerate}
\nomenclature{$\mathbf{H}_0^\mathbb{N}$}{Null hypothesis for testing the NASA experiment}
\nomenclature{$\mathbf{H}_A^\mathbb{N}$}{Alternate hypothesis for testing power reduction in the NASA experiment}

\subsubsection{Regression-based modeling}~\\
\label{ssec:nasa_regression}
\indent Given the statistical significance, is of interest to predict the experiment phase power consumption based on a model. To model the hourly power consumption of an average participant, we employ an autoregressive model with an exogenous input consisting of the average screentime associated with the dashboard during the past hour. In place of specifying the start time ($t_0$) and end time ($t_f$) arguments, let the single argument $h \in \{1,...,24\}$ specify the hour of the day $d$. Therefore, we can write the experiment and baseline hourly power consumption of the $i^{th}$ participant during hour $h$ on day $d$ as $Y_{\mathcal{P}3^\mathbb{N}_i}(d,h)$ and $Y_{\mathcal{P}1^\mathbb{N}_i}(d,h)$, respectively. Similarly, the intervention variable can be written as $X^A_{\mathcal{P}3^\mathbb{N}_i}(d,h)$. Also, let '$:$' denote an index representing the sample statistic when used in place of '$i$', the index corresponding to the $i^{th}$ participant. Instead of explicitly modeling the experiment hourly power consumption of an average participant $Y_{\mathcal{P}3^\mathbb{N}_:}(d,h)$, we model the \emph{difference between the averaged baseline and experiment responses}, $\mu_{\mathcal{P}1^\mathbb{N}_:}(d,h)-Y_{\mathcal{P}3^\mathbb{N}_:}(d,h)$\footnote{The difference is constructed by considering the baseline week day $d \in \mathbb{D}_W$ corresponding to the experiment day $d \in D_3^\mathbb{N}$.\label{regression_differential}}. The paired difference mitigates subjective variation due to individual energy consumption arising from varied energy preferences or work schedules, thereby enabling better prediction. Let this mean differential response be represented by $\Delta Y_{\mathbb{N}_{:}}(d,h)$. \added{To model this response, we considered various model families differing in structural complexity. These included timeseries regression using linear, polynomial, and logarithmic functions; neural networks; kernel regression; and Gaussian process regression. It was observed that increasing the model complexity did not necessarily translate to tangible improvements in performance (< 5\%). Based on the evaluation, the autoregressive model structure with exogenous inputs (ARX) shown below is chosen.} \deleted{Now, we construct an AutoRegression model with eXogenous inputs (ARX) as follows}:
\begin{equation}
\label{eqn:arx_nasa}
\Delta Y_{\mathbb{N}_{:}}(d,h) = \alpha^\mathbb{N} + \beta^\mathbb{N}\Delta Y_{\mathbb{N}_{:}}(d,h-1) + \delta^\mathbb{N} X^A_{\mathcal{P}3^\mathbb{N}_:}(d,h-1) + \epsilon^\mathbb{N}(d,h)
\end{equation}
where, $\epsilon^\mathbb{N}(d,h)$ represents the error process following a Gaussian distribution $\mathcal{N}(0,\sigma^\mathbb{N}_\epsilon)$. The introduction of the lagged variable $\Delta Y_{\mathbb{N}_{:}}(d,h-1)$ is instrumental in weakening the residual serial correlation and thus mitigates systematic factors in the error process as shown in Figure \ref{FIG:plugload_nasa_autocorr_lags}. It depicts the impact of adding time-lagged dependent variables on the serial correlation of the residuals. It is evident that the first order lag significantly reduces the correlation and the introduction of further lags do not contribute toward reducing the correlation further. From an experiment perspective, the time-lagged dependent variable enables us to account for changes between experiment conditions with respect to the baseline conditions. For example, any change in workload between the baseline and the experiment conditions can be captured by the introduction of the time-lagged dependent term in the model. This allows us to strengthen the assumption that the residuals corresponding to consecutive hours are a result of random factors and hence uncorrelated given the inputs. Given the justified model in Equation \ref{eqn:arx_nasa} we train and test it against the data collected from experiment $\mathbb{E_N}$ and the results are discussed in Section \ref{ssec:nasa_regression_results}.

\subsection{Data analysis for experiment $\mathbb{E_C}$ at CMU SV}
\label{ssec:cmu_analysis}
The experiment $\mathbb{E_C}$ was conducted in four phases: A baseline phase and three experiment phases. The experiment phases $\mathcal{P}2^\mathbb{C}$, $\mathcal{P}3^\mathbb{C}$, and $\mathcal{P}4^\mathbb{C}$ consisted of interventions in the form of \emph{incentives}, \emph{feedback}, and \emph{both incentives and feedback}, respectively. Similar to the experiment $\mathbb{E_N}$, during phase $\mathcal{P}k^\mathbb{C}\ (k \in \{1,2,3,4\})$, let the instantaneous power consumption of the $i^{th}$ participant $\mathcal{P}k^\mathbb{C}_i$ on the day $d\ (\in D_k^\mathbb{C})$ at instant $t$ be denoted by $y_{\mathcal{P}k^\mathbb{C}_i}(d,t)$, and let the average power consumption during $[t_0,t_f)$ be denoted by $\bar{y}_{\mathcal{P}k^\mathbb{C}_i}(d,t_0,t_f)$. Also, let the incentive and feedback provided during the time interval $[t_0,t_f)$ for the respective phases be denoted by $x^I_{\mathcal{P}k^\mathbb{C}_i}(d,t_0,t_f),\ (k \in \{3,4\})$ and $x^A_{\mathcal{P}k^\mathbb{C}_i}(d,t_0,t_f)$\footnote{The scoreboard feature of the dashboard was made visible to the participants even during the incentive phase $(k=2)$ to let them monitor their position in competing for the incentive. Hence, in $\mathbb{E_C}$, the screentime variable is applicable to all the experiment phases $\mathcal{P}k^\mathbb{C}\ (k \in \{2,3,4\})$. \label{leaderboard_feature}}, respectively.
\nomenclature{$y_{\mathcal{P}1^\mathbb{C}_i}(t_0,t_f)$}{Baseline response of the $i^{th}$ participant during the time interval $[t_0,t_f)$ in the CMU experiment}
\nomenclature{$y_{\mathcal{P}2^\mathbb{C}_i}(t_0,t_f)$}{Response of the $i^{th}$ participant during the time interval $[t_0,t_f)$ in the CMU incentive experiment}
\nomenclature{$y_{\mathcal{P}3^\mathbb{C}_i}(t_0,t_f)$}{Response of the $i^{th}$ participant during the time interval $[t_0,t_f)$ in the CMU feedback experiment}
\nomenclature{$y_{\mathcal{P}4^\mathbb{C}_i}(t_0,t_f)$}{Response of the $i^{th}$ participant during the time interval $[t_0,t_f)$ in the CMU \emph{both feedback and incentive} experiment}
\nomenclature{$x^A_{\mathcal{P}2^\mathbb{C}_i}(t_0,t_f)$}{screentime spent by the $i^{th}$ participant on the scoreboard during the time interval $[t_0,t_f)$ in the CMU incentive experiment}
\nomenclature{$x^I_{\mathcal{P}2^\mathbb{C}_i}(t_0,t_f)$}{Incentive input to the $i^{th}$ participant during the time interval $[t_0,t_f)$ in the CMU incentive experiment}
\nomenclature{$x^A_{\mathcal{P}3^\mathbb{C}_i}(t_0,t_f)$}{screentime spent by the $i^{th}$ participant on the dashboard during the time interval $[t_0,t_f)$ in the CMU feedback experiment}
\nomenclature{$x^A_{\mathcal{P}4^\mathbb{C}_i}(t_0,t_f)$}{screentime spent by the $i^{th}$ participant on the dashboard during the time interval $[t_0,t_f)$ in the CMU \emph{both feedback and incentive} experiment}
\nomenclature{$x^I_{\mathcal{P}4^\mathbb{C}_i}(t_0,t_f)$}{Incentive input to the $i^{th}$ participant during the time interval $[t_0,t_f)$ in the CMU \emph{both feedback and incentive} experiment}

For inference, we regard the observations as the realizations of a random sample from the occupant population. Similar to the experiment $\mathbb{E_N}$, we use $Y_{\mathcal{P}k^\mathbb{C}_i}(d,t)$, $Y_{\mathcal{P}k^\mathbb{C}_i}(d,t_0,t_f)$, $X^A_{\mathcal{P}k^\mathbb{C}_i}(d,t_0,t_f)$, and $X^I_{\mathcal{P}k^\mathbb{C}_i}(d,t_0,t_f)$ to denote random variables corresponding to the $i^{th}$ participant and the experiment day $d \in D_k^\mathbb{C}$. Thus, the random variables pertaining to the response of the $i^{th}$ participant on day $d$ during the time interval $[t_0,t_f)$ corresponding to each of the phases $\mathcal{P}1^\mathbb{C}$, $\mathcal{P}2^\mathbb{C}$, $\mathcal{P}3^\mathbb{C}$, and $\mathcal{P}4^\mathbb{C}$, by convention, become $Y_{\mathcal{P}1^\mathbb{C}_i}(d,t_0,t_f)$, $Y_{\mathcal{P}2^\mathbb{C}_i}(d,t_0,t_f)$, $Y_{\mathcal{P}3^\mathbb{C}_i}(d,t_0,t_f)$, and $Y_{\mathcal{P}4^\mathbb{C}_i}(d,t_0,t_f)$, respectively. Similarly, the corresponding random variables representing the interventions during each of the three experiment phases $\mathcal{P}2^\mathbb{C}$, $\mathcal{P}3^\mathbb{C}$, and $\mathcal{P}4^\mathbb{C}$ become $\big(X^A_{\mathcal{P}2^\mathbb{C}_i}(d,t_0,t_f),X^I_{\mathcal{P}2^\mathbb{C}_i}(d,t_0,t_f)\big)$, $\big(X^A_{\mathcal{P}3^\mathbb{C}_i}(d,t_0,t_f)\big)$, and $\big(X^A_{\mathcal{P}4^\mathbb{C}_i}(d,t_0,t_f),X^I_{\mathcal{P}4^\mathbb{C}_i}(d,t_0,t_f)\big)$, respectively.
\nomenclature{$Y_{\mathcal{P}1^\mathbb{C}_i}(t_0,t_f)$}{Random variable representing the baseline response of the $i^{th}$ participant during the time interval $[t_0,t_f)$ in the CMU experiment}
\nomenclature{$Y_{\mathcal{P}2^\mathbb{C}_i}(t_0,t_f)$}{Random variable representing the response of the $i^{th}$ participant during the time interval $[t_0,t_f)$ in the CMU incentive experiment}
\nomenclature{$Y_{\mathcal{P}3^\mathbb{C}_i}(t_0,t_f)$}{Random variable representing the response of the $i^{th}$ participant during the time interval $[t_0,t_f)$ in the CMU dashboard feedback experiment}
\nomenclature{$Y_{\mathcal{P}4^\mathbb{C}_i}(t_0,t_f)$}{Random variable representing the response of the $i^{th}$ participant during the time interval $[t_0,t_f)$ in the CMU \emph{both feedback and incentive} experiment}
\nomenclature{$x^A_{\mathcal{P}2^\mathbb{C}_i}(t_0,t_f)$}{Random variable representing the screentime spent by the $i^{th}$ participant on the scoreboard during the time interval $[t_0,t_f)$ in the CMU incentive experiment}
\nomenclature{$x^I_{\mathcal{P}2^\mathbb{C}_i}(t_0,t_f)$}{Random variable representing the incentive input of the $i^{th}$ participant during the time interval $[t_0,t_f)$ in the CMU incentive experiment}
\nomenclature{$x^A_{\mathcal{P}3^\mathbb{C}_i}(t_0,t_f)$}{Random variable representing the screentime spent by the $i^{th}$ participant on the dashboard during the time interval $[t_0,t_f)$ in the CMU dashboard feedback experiment}
\nomenclature{$x^A_{\mathcal{P}4^\mathbb{C}_i}(t_0,t_f)$}{Random variable representing the screentime spent by the $i^{th}$ participant on the dashboard during the time interval $[t_0,t_f)$ in the CMU \emph{both incentive and dashboard} experiment}
\nomenclature{$x^I_{\mathcal{P}4^\mathbb{C}_i}(t_0,t_f)$}{Random variable representing the incentive input of the $i^{th}$ participant during the time interval $[t_0,t_f)$ in the CMU \emph{both feedback and incentive} experiment}
\subsubsection{Statistical assumptions}~\\
\label{ssec:cmu_stat_assumptions}
\indent Given the baseline and experimental conditions, let the day of the week d ($\in \mathbb{D}_W$) statistics corresponding to $Y_{\mathcal{P}1^\mathbb{C}_i}(d,t_0,t_f)$, $Y_{\mathcal{P}2^\mathbb{C}_i}(d,t_0,t_f)$, $Y_{\mathcal{P}3^\mathbb{C}_i}(d,t_0,t_f)$, and $Y_{\mathcal{P}4^\mathbb{C}_i}(d,t_0,t_f)$ by $\mu_{\mathcal{P}1^\mathbb{C}_i}(d,t_0,t_f)$, $\mu_{\mathcal{P}2^\mathbb{C}_i}(d,t_0,t_f)$, $\mu_{\mathcal{P}3^\mathbb{C}_i}(d,t_0,t_f)$, and $\mu_{\mathcal{P}4^\mathbb{C}_i}(d,t_0,t_f)$. 
For performing inference, similar to experiment $\mathbb{E_N}$, we consider the sample constituted by the differences in the daily-averaged experiment response and the corresponding baseline response, sampled across participants and days of the week. For a day $d \in \mathbb{D}_W$ in the experiment phase $j$ ($\in \{2,3,4\}$), let this averaged response differential for the $i^{th}$ participant be represented by the random sample $\mu_{\mathcal{P}1^\mathbb{C}_i}(d,0,86400)-\mu_{\mathcal{P}j^\mathbb{C}_i}(d,0,86400)$. The sample elements are assumed to be independent across the days of the week and the participant pool. Given the matched pairs design similar to Section \ref{ssec:nasa_stat_assumptions}, any inferences from the differential sample can be attributed to the intervention(s) administered during $\mathcal{P}j^\mathbb{C}$ instead of the nuisance factors stated in Section \ref{expt_principles}.
\subsubsection{Hypothesis testing and confidence interval estimation}~\\
\label{ssec:cmu_hypothesis_testing}
\indent Given the assumptions about the population consisting of the differential responses we employ a paired difference t-test to draw inferences about the underlying population. For each of the experiment phases $\mathcal{P}j^\mathbb{C},\ j \in \{2,3,4\}$, hypothesis tests and confidence interval estimation are performed on the mean of the differential response $\mu_{\mathcal{P}1^\mathbb{C}_i}(d,0,86400)-\mu_{\mathcal{P}j^\mathbb{C}_i}(d,0,86400)$. Thus, the null and alternate hypothesis for $\mathcal{P}j^\mathbb{C}$ become:
\begin{enumerate}
\item $\mathbf{H}_0^{j\mathbb{C}}$: $\mu_{\mathcal{P}1^\mathbb{C}_i}(d,0,86400)-\mu_{\mathcal{P}j^\mathbb{C}_i}(d,0,86400)$ is sampled from a population with zero mean
\item $\mathbf{H}_A^{j\mathbb{C}}$: $\mu_{\mathcal{P}1^\mathbb{C}_i}(d,0,86400)-\mu_{\mathcal{P}j^\mathbb{C}_i}(d,0,86400)$ is sampled from a population with non-zero mean
\end{enumerate}

\subsubsection{Regression-based modeling}
\begin{figure}[ht!]
     \begin{center}
        \subfigure[Experiment $\mathbb{E_N}$: Lag 1 correlation coefficient of the residual process vs number of lagged dependents]{%
            \centering
            \includegraphics[width=0.4\textwidth,keepaspectratio]{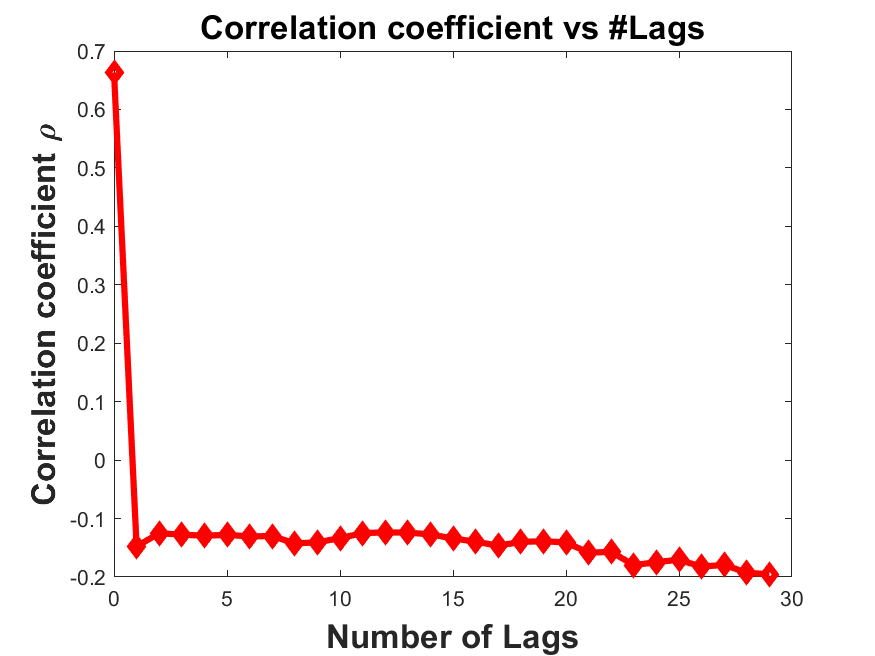}
            \label{FIG:plugload_nasa_autocorr_lags}
        }\hfill
        \subfigure[Experiment $\mathbb{E_C}$: Lag 1 correlation coefficient of the residual process vs number of lagged dependents]{%
            \centering
            \includegraphics[width=0.4\textwidth,keepaspectratio]{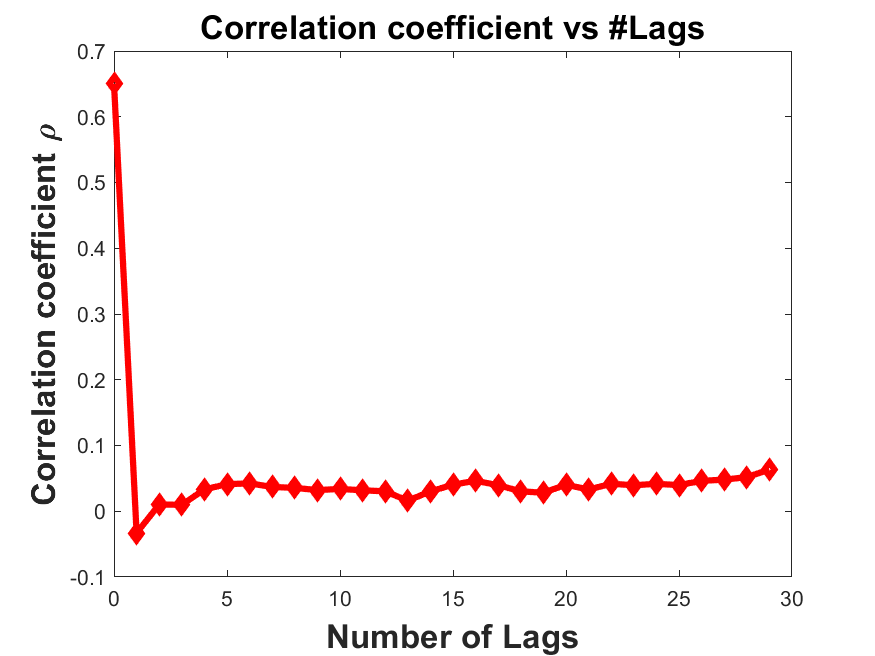}
            \label{FIG:plugload_cmu_autocorr_lags}
        }\\ 
        \subfigure[Residual analysis of the autoregressive model corresponding to experiment $\mathbb{E_N}$]{%
            \centering
            \includegraphics[width=0.4\textwidth,keepaspectratio]{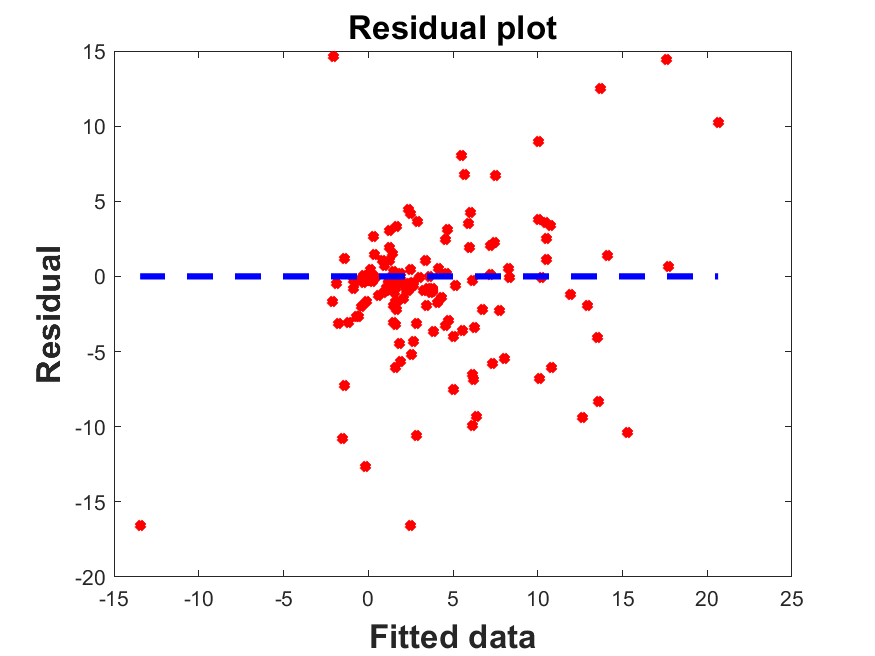}
            \label{FIG:nasa_residuals}
        }\hfill
        \subfigure[Residual analysis of the autoregressive model corresponding to experiment $\mathbb{E_C}$]{%
            \centering
            \includegraphics[width=0.4\textwidth,keepaspectratio]{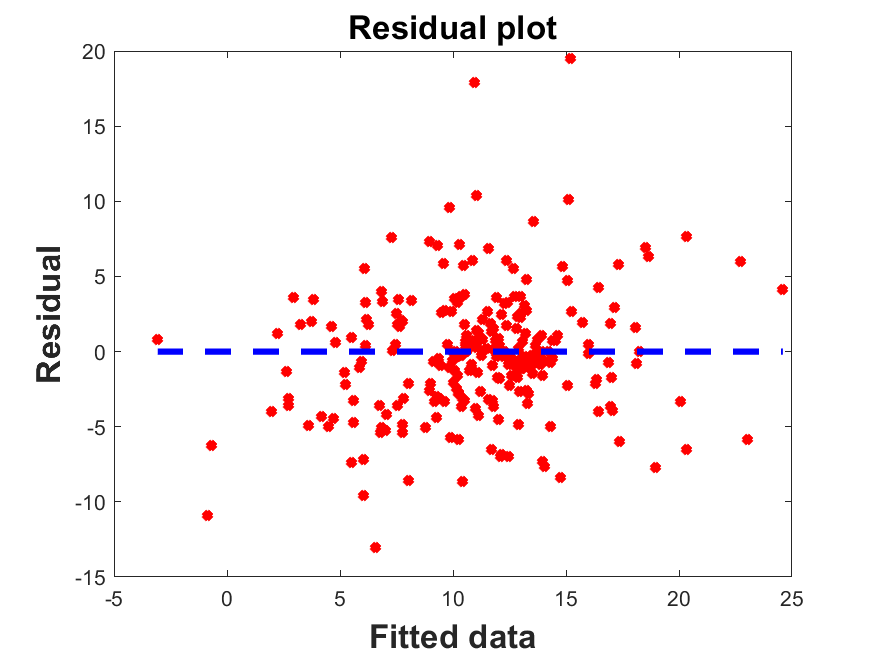}
            \label{FIG:cmu_residuals}
        }\\ 
    \end{center}
    \caption{Statistical modeling results}%
   \label{FIG:statresults_modeling}
\end{figure}

~\\
\label{ssec:cmu_regression}
\indent Given the interval estimates, we are interested in a predictive model akin to the one in Section \ref{ssec:nasa_regression}. We employ a similar notation here. In case of experiment $\mathbb{E_N}$, dashboard feedback was the only intervention used and hence screentime was the only exogenous variable considered. However, in experiment $\mathbb{E_C}$, each phase consists of either an incentive intervention (in phase $\mathcal{P}2^\mathbb{C}$) or a dashboard feedback intervention (in phase $\mathcal{P}3^\mathbb{C}$) or both (in phase $\mathcal{P}4^\mathbb{C}$). For modeling purposes, we note that each observation in phase $k \in \{2,3,4\}$ can have a non-negative value for each of the intervention variables $x^A_{\mathcal{P}k^\mathbb{C}_:}(d,h)$\footref{leaderboard_feature} and $x^I_{\mathcal{P}k^\mathbb{C}_:}(d,h)$, thereby simultaneously accommodating both exogenous inputs into the model structure. Further, let the baseline and experiment hourly power consumption of the $i^{th}$ participant during hour $h$ on the experiment day $d$ be denoted by $Y_{\mathcal{P}1^\mathbb{C}_{:}}(d,h)$ and $Y_{\mathcal{P}e^\mathbb{C}_{:}}(d,h)$, respectively. We then model the mean differential response $\Delta Y_{\mathbb{C}_{:}}(d,h) := \mu_{\mathcal{P}1^\mathbb{C}_{:}}(d,h)-Y_{\mathcal{P}e^\mathbb{C}_{:}}(d,h)$\footref{regression_differential} by an ARX model consisting of screentime, the incentive, and their interaction as the inputs. \added{Similar to the model selection process described in Section \ref{ssec:nasa_regression}, we evaluated various model structures prior to selecting the following ARX model:}. \deleted{Thus, we have:}
\begin{equation}
\label{eqn:arx_cmu}
\Delta Y_{\mathbb{C}_{:}}(d,h) = \alpha^\mathbb{C} + \beta^\mathbb{C}\Delta Y_{\mathbb{C}_{:}}(d,h-1) + \gamma^\mathbb{C}x^I_{\mathcal{P}e^\mathbb{C}_:}(d,h-1) + \delta^\mathbb{C}x^A_{\mathcal{P}e^\mathbb{C}_:}(d,h-1) + \tau^\mathbb{C}x^I_{\mathcal{P}e^\mathbb{C}_:}(d,h-1)x^A_{\mathcal{P}e^\mathbb{C}_:}(d,h-1) + \epsilon^\mathbb{C}(d,h)
\end{equation}
where, $\epsilon^\mathbb{C}(d,h)$ represents the error process following a Gaussian distribution $\mathcal{N}(0,\sigma^\mathbb{C}_\epsilon)$. The introduction of the lagged dependent term $\Delta Y_{\mathbb{C}_{:}}(d,h-1)$ is instrumental in weakening the residual serial correlation. 
Figure \ref{FIG:plugload_cmu_autocorr_lags} depicts the relationship between the number of added lags and the residual correlation. It is evident that the additional lags do not add further systematic information about the predicted variable and hence do not significantly contribute toward weakening the residual serial correlation. From an experiment standpoint, these lags capture the change in experiment conditions as compared to the baseline conditions, thereby strengthening the assumption that the residuals corresponding to consecutive hours are uncorrelated given the model inputs. The justified model in Equation \ref{eqn:arx_cmu} is trained and tested against the data collected from experiment $\mathbb{E_C}$, and the results are discussed in Section \ref{ssec:cmu_regression_results}.

\begin{figure}[ht!]
     \begin{center}
        \subfigure[Experiment $\mathbb{E_N}$: Statistical summary of the data from phase $\mathcal{P}3^\mathbb{N}$]{%
            \centering
            \includegraphics[width=0.4\textwidth,keepaspectratio]{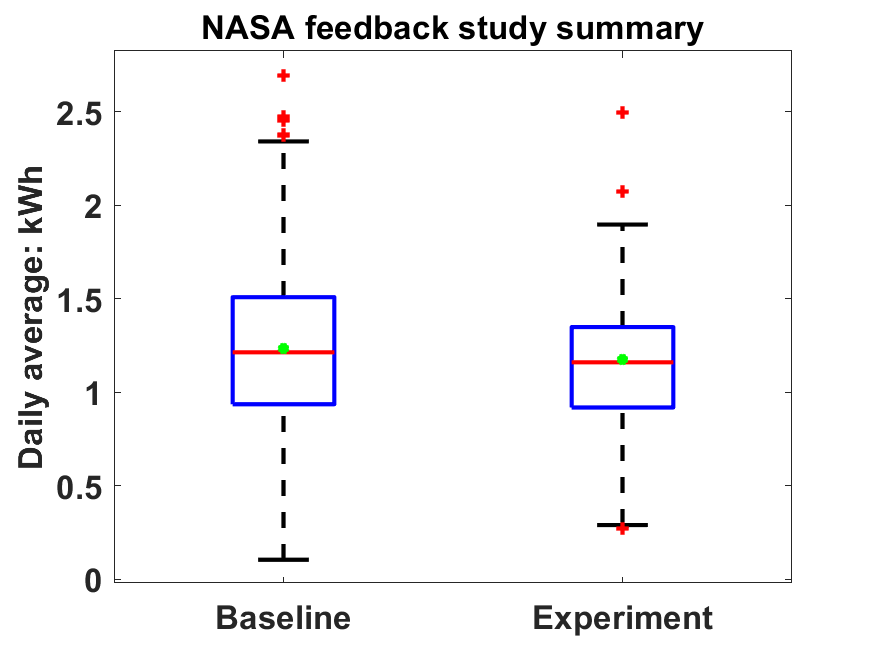}
            \label{FIG:stat_summary_nasa_g1_g3}
        }\hfill
        \subfigure[Experiment $\mathbb{E_C}$: Statistical summary of the data from phase $\mathcal{P}2^\mathbb{C}$]{%
            \centering
            \includegraphics[width=0.4\textwidth,keepaspectratio]{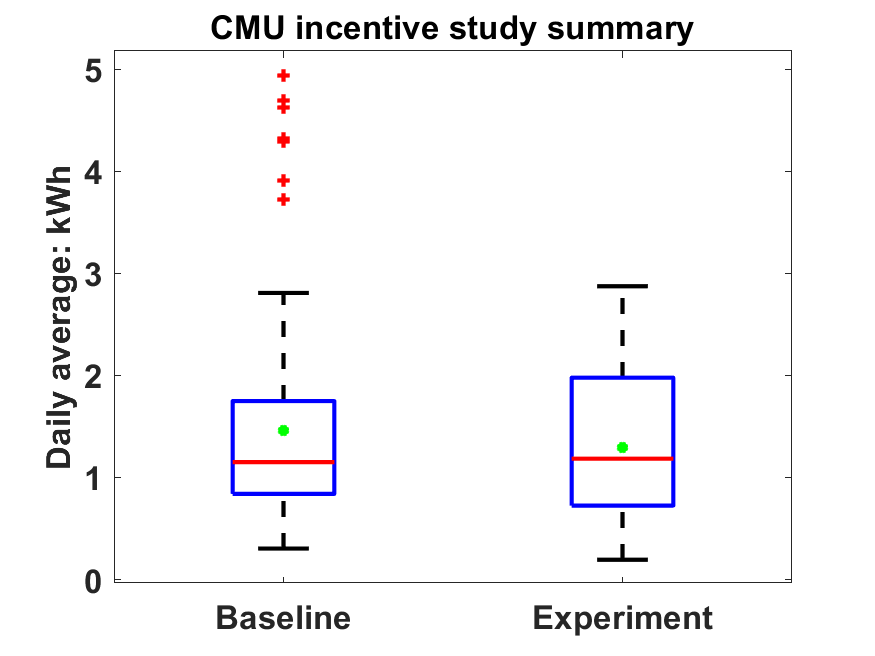}
            \label{FIG:stat_summary_cmu_g1_g2}
        }\\ 
        \subfigure[Experiment $\mathbb{E_C}$: Statistical summary of the data from phase $\mathcal{P}3^\mathbb{C}$]{%
            \centering
            \includegraphics[width=0.4\textwidth,keepaspectratio]{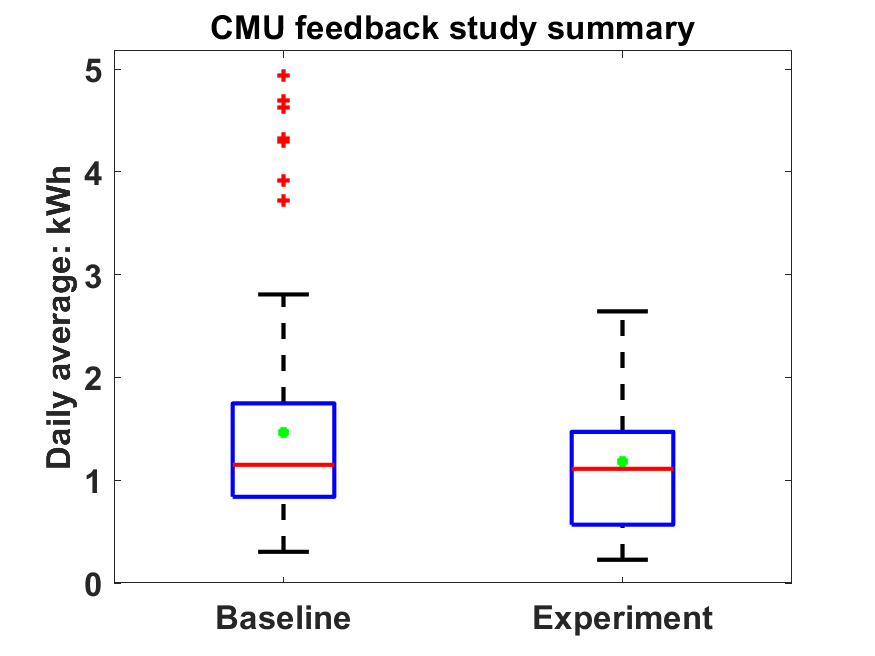}
            \label{FIG:stat_summary_cmu_g1_g3}
        }\hfill
        \subfigure[Experiment $\mathbb{E_C}$: Statistical summary of the data from phase $\mathcal{P}4^\mathbb{C}$]{%
            \centering
            \includegraphics[width=0.4\textwidth,keepaspectratio]{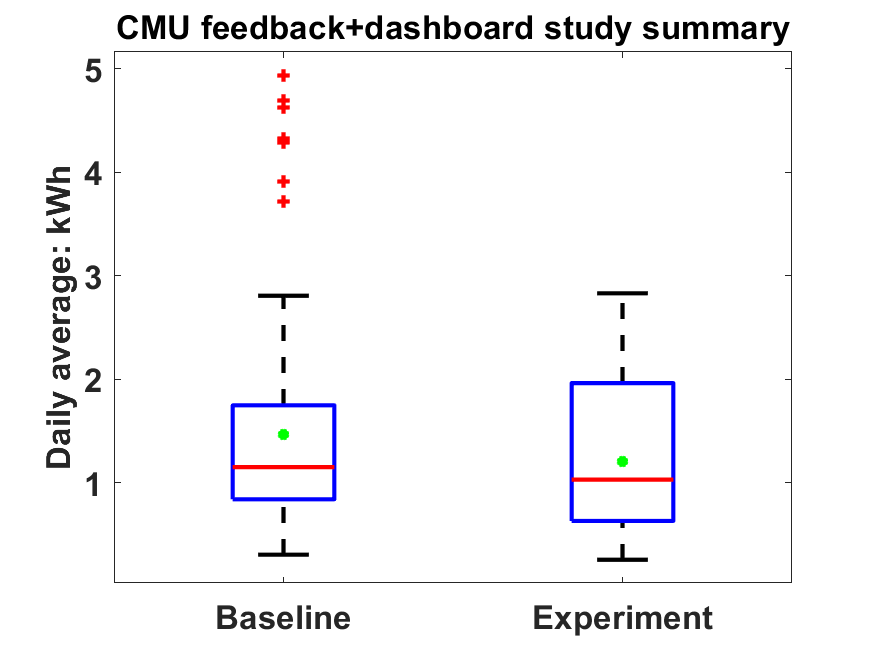}
            \label{FIG:stat_summary_cmu_g1_g4}
        }\\ 
    \end{center}
    \caption{Statistical summary of the data from experiments}%
   \label{FIG:statresults_summary}
\end{figure}
\begin{figure}[ht!]
     \begin{center}
        \subfigure[Power prediction on the test set based on the data from experiment $\mathbb{E_N}$]{%
            \centering
            \includegraphics[width=\textwidth,keepaspectratio]{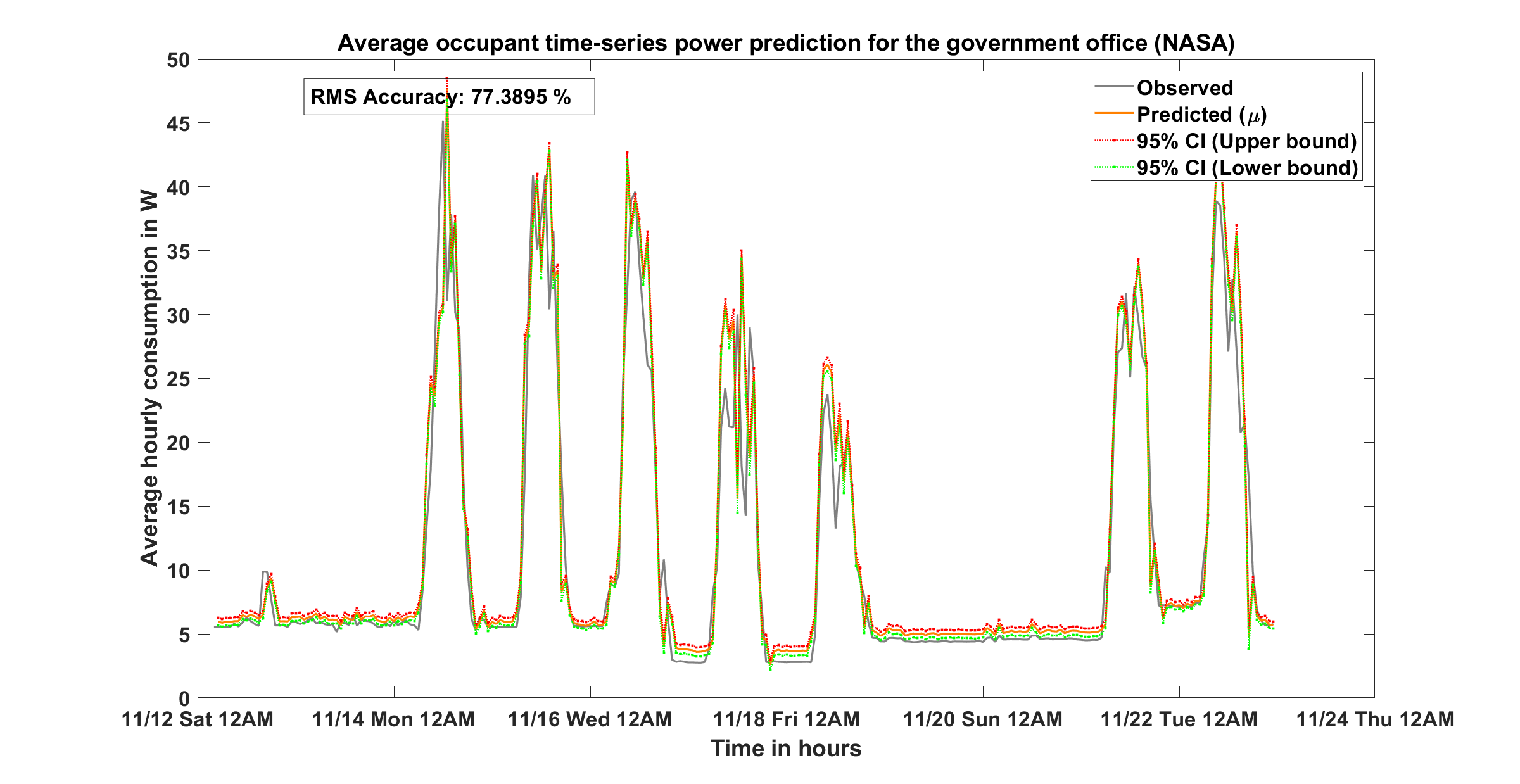}
            \label{FIG:nasa_ts_power_prediction}
        }\\ 
        \subfigure[Power prediction on the test set based on the data from experiment $\mathbb{E_C}$]{%
           \centering
            \includegraphics[width=\textwidth,keepaspectratio]{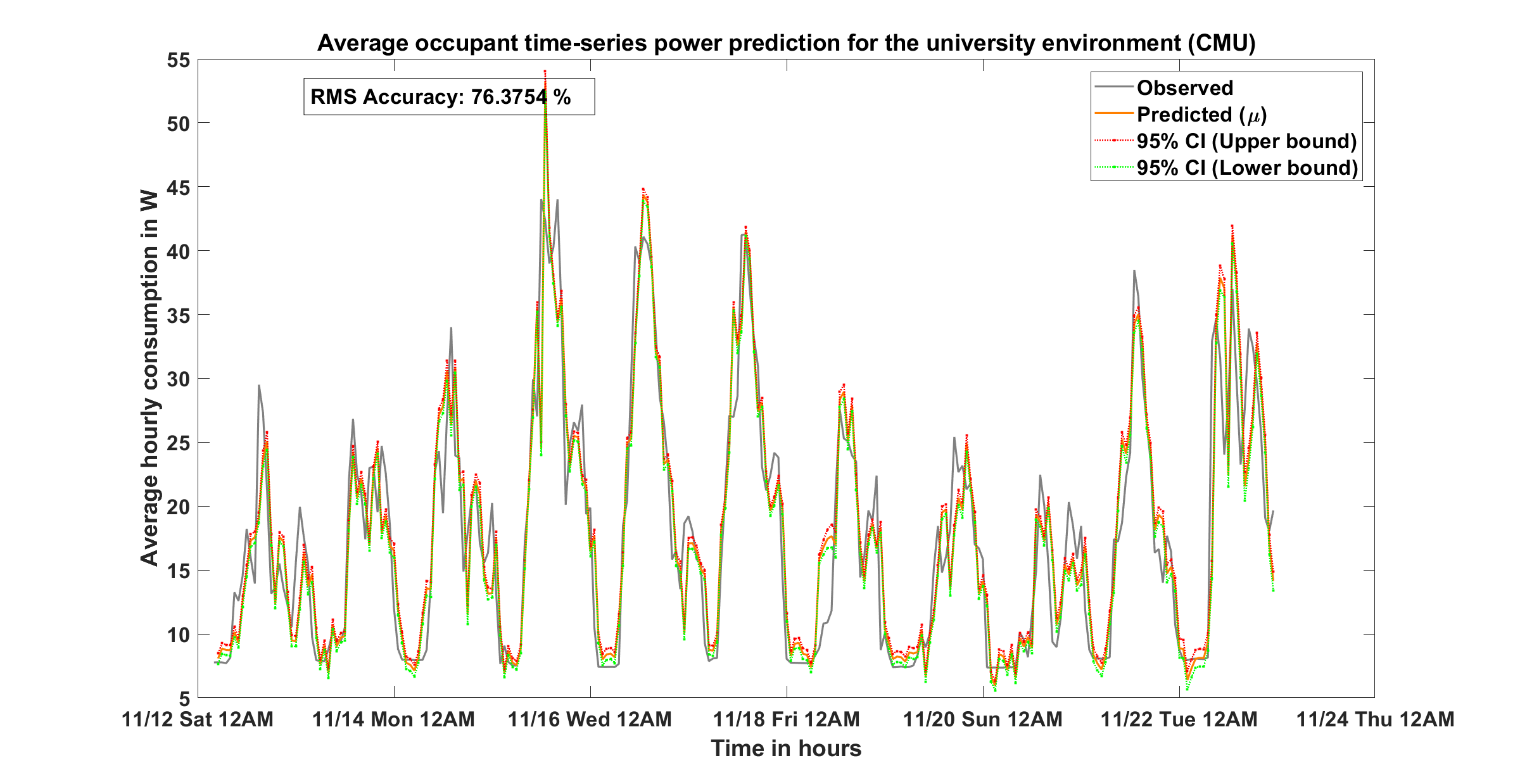}
            \label{FIG:cmu_ts_power_prediction}
        }%
    \end{center}
    \caption{Performance of predictive models}%
   \label{FIG:regressionresults}
\end{figure} 
\section{Results and discussion}
\label{sec:results_and_discussion}
The results from hypotheses testing, interval estimation, and regression modeling for the experiments $\mathbb{E_N}$ and $\mathbb{E_C}$ are presented in Sections \ref{ssec:nasa_results} and \ref{ssec:cmu_results}, respectively. These results are discussed in Section \ref{ssec:discussion}.
\subsection{Results from experiment $\mathbb{E_N}$}
\label{ssec:nasa_results}
As per the design formulated in Section \ref{ssec:nasa_analysis}, we present the results from the government office experiment below.
\subsubsection{Hypothesis testing and confidence interval estimation:}
\label{ssec:nasa_ht_results}
Consider the hypotheses in Section \ref{ssec:nasa_hyp_test}. The mean baseline power consumption and the mean feedback phase power consumption are 51.51 W and 48.86 W, respectively. The t-statistic is found to be t(86)=3.64, and the corresponding p-value is $p=4.61\times10^{-4}$. Therefore, the evidence against the null hypothesis is statistically significant at $\alpha=0.05$. Further, we note that the 95\% confidence interval of the mean of the differential sample $\mu_{\mathcal{P}1^\mathbb{N}_i}(d,0,86400)-\mu_{\mathcal{P}3^\mathbb{N}_i}(d,0,86400)$ is [2.22, 7.57] W. Thus, we conclude that the mean power consumption during the feedback phase $\mathcal{P}3^\mathbb{N}$ is (statistically) significantly \emph{less} than that of the baseline phase $\mathcal{P}1^\mathbb{N}$. The summary data and test results are presented in Figure \ref{FIG:stat_summary_nasa_g1_g3} and Table \ref{table:nasa_ht_interval_results}, respectively.
\subsubsection{Autoregressive model:}
\label{ssec:nasa_regression_results}
Based on the model designed in Section \ref{ssec:nasa_regression}, we estimate the parameters on the training set using $70\%$ of the data. These parameters are estimated by Ordinary Least Squares (OLS) and are provided in Table \ref{table:nasa_regression_results}. The coefficients $\beta^\mathbb{N}=0.8042$ and $\delta^\mathbb{N}=0.0019$ represent the effect of last hour power differential in Watts and the effect of change in screentime in seconds, respectively. While both coefficients are positively correlated with the predicted variable, the difference in their magnitudes stems from the differences in the units of the predictors and their predictive contributions. The intercept $\alpha^\mathbb{N}=-0.0298$ represents the constant residual effect after adjusting for the predictor variables given the zero-mean gaussian process $\epsilon^\mathbb{N}(d,h)$. The model performance on the test dataset is shown in Figure \ref{FIG:nasa_ts_power_prediction}. The RMS residuals on the training and test sets are 3.51W and 3.53W, respectively. The RMS accuracy of the model on the test set is $\approx 77.39\%$.

\subsection{Results from experiment $\mathbb{E_C}$}
\label{ssec:cmu_results}
As per the design formulated in Section \ref{ssec:cmu_analysis}, we present the results from the university experiment below.
\subsubsection{Hypothesis testing and confidence interval estimation:}
\label{ssec:cmu_ht_results}
Similar to experiment $\mathbb{E_N}$, we consider the hypotheses for each of the intervention phases as outlined in Section \ref{ssec:cmu_hypothesis_testing}.
\paragraph{Incentive phase ($\mathcal{P}2$):}We find the mean baseline power consumption and the mean incentive phase power consumption to be 60.90 W and 53.91 W, respectively. The t-statistic is found to be $t(74)=1.62$, and the corresponding p-value is $p=0.11$. The $95\%$ confidence interval of the mean of the differential response $\mu_{\mathcal{P}1^\mathbb{C}_i}(d,0,86400)-\mu_{\mathcal{P}2^\mathbb{C}_i}(d,0,86400)$ is [-1.84, 17.63] W. Thus, the mean power consumption during the incentive phase is not statistically different from the baseline power consumption at $\alpha=0.05$. The summary data and test results are presented in Figure \ref{FIG:stat_summary_cmu_g1_g2} and Table \ref{table:cmu_ht_interval_results}, respectively.

\paragraph{Feedback phase ($\mathcal{P}3$):}We find that the mean baseline power consumption and the mean feedback phase power consumption are 60.90 W and 49.27 W, respectively. The t-statistic is found to be $t(75)=2.26$, and the corresponding p-value is $p=0.03$. The $95\%$ confidence interval of the mean of the differential response $\mu_{\mathcal{P}1^\mathbb{C}_i}(d,0,86400)-\mu_{\mathcal{P}3^\mathbb{C}_i}(d,0,86400)$ is [1.58,24.82] W. Thus, the mean power consumption during the feedback phase is statistically significantly less than that of the baseline phase at $\alpha=0.05$. The summary data and test results are presented in Figure \ref{FIG:stat_summary_cmu_g1_g3} and Table \ref{table:cmu_ht_interval_results}, respectively.

\paragraph{Feedback and Incentive phase ($\mathcal{P}4$):}We find that the mean baseline power consumption and the mean feedback \& incentive phase power consumption are 60.90 W and 50.33 W, respectively. The t-statistic is found to be $t(67)=2.30$, and the corresponding p-value of $p=0.02$. The $95\%$ confidence interval of the mean of the differential response $\mu_{\mathcal{P}1^\mathbb{C}_i}(d,0,86400)-\mu_{\mathcal{P}4^\mathbb{C}_i}(d,0,86400)$ is [1.96,27.63] W. Thus, the mean power consumption during the feedback \& incentive phase is statistically significantly less than that of the baseline phase at $\alpha=0.05$. The  summary data and test results are presented in Figure \ref{FIG:stat_summary_cmu_g1_g4} and Table \ref{table:cmu_ht_interval_results}, respectively.

\begin{table}
\parbox{.4\linewidth}{
\centering
\resizebox{0.7\linewidth}{!}{
    \begin{tabular}{| l | l | l |}
    \hline
    \diagbox{\textbf{Variable}}{\textbf{Phase}} & $\mathcal{P}3$ \\ \hline
    Sample mean (W) & 2.65 \\ \hline
    df & 86 \\ \hline
    t-statistic & 3.64 \\ \hline
    p-value & $4\times 10^{-4}$ \\ \hline
    Hypothesis chosen & $\mathbf{H}_A^\mathbb{N}$ \\ [2pt] \hline
    95\% C.I. (W) & [2.22, 7.57] \\ \hline
    95\% C.I. ($\%$) & [4.32, 14.71] \\
    \hline
    \end{tabular}}
    \caption{Hypothesis test results and interval estimates for $\mathbb{E}_N$}
    \label{table:nasa_ht_interval_results}
}
\hfill
\parbox{.7\linewidth}{
\centering
\resizebox{0.7\linewidth}{!}{
    \begin{tabular}{| l | l | l | l | l |}
    \hline
    \diagbox{\textbf{Variable}}{\textbf{Phase}} & $\mathcal{P}2$ & $\mathcal{P}3$ & $\mathcal{P}4$ \\ \hline
	Sample mean (W) & 7.18 & 11.82 & 10.76 \\ \hline
    df & 74 & 75 & 67 \\ \hline
    t-statistic & 1.62 & 2.26 & 2.30 \\ \hline
    p-value & 0.11 & 0.03 & 0.02 \\ \hline
    Hypothesis chosen & $\mathbf{H}_0^{2\mathbb{C}}$ & $\mathbf{H}_A^{3\mathbb{C}}$ & $\mathbf{H}_A^{4\mathbb{C}}$ \\ [2pt] \hline
    95\% CI (W) & [-1.84, 17.63] & [1.58, 24.82] & [1.96, 27.63] \\ \hline
    95\% CI ($\%$) & [-3.01, 28.87] & [2.59, 40.63] & [3.21, 45.24] \\
    \hline
	\end{tabular}}
    \caption{Hypothesis test results and interval estimates for $\mathbb{E}_C$}
    \label{table:cmu_ht_interval_results}
}
\end{table}

\subsubsection{Autoregressive model:}
\label{ssec:cmu_regression_results}
Consider the the model designed in Section \ref{ssec:cmu_regression}. We estimate the parameters with $70\%$ data via OLS similar to that of Section \ref{ssec:nasa_regression_results}. These estimates are listed in table \ref{table:cmu_regression_results}. The coefficient $\beta^\mathbb{C}=0.7679$ represents the effect of the last hour power differential in Watts. Similarly, the predictive effects of the incentive in $\$$, the dashboard feedback in seconds, and their interaction in $\$$-seconds are represented by $\gamma^\mathbb{C}=-0.0048$, $\delta^\mathbb{C}=0.0045$, and $\tau^\mathbb{C}=0.0001$, respectively. The switch of signs between the incentive and the interaction coefficients suggests a crossover interaction between the feedback and the incentive for purposes of prediction. The intercept term $\alpha^\mathbb{C}=2.4667$ represents the constant residual effect after accounting for the predictors given the zero-mean gaussian process $\epsilon^\mathbb{C}(d,h)$. The model performance on the test dataset is shown in Figure $\ref{FIG:cmu_ts_power_prediction}$. The RMS residuals on the training and test sets are 3.95W and 4.09W, respectively. The RMS accuracy of the model on the test set is $\approx 76.38\%$.
\begin{table}
\parbox{.45\linewidth}{
\centering
\resizebox{0.93\linewidth}{!}{
    \begin{tabular}{| l | l | l | l | l |}
    \hline
    \diagbox{\textbf{Property}}{\textbf{Coeff.}} & $\alpha^\mathbb{N}$ & $\beta^\mathbb{N}$ & $\delta^\mathbb{N}$ & $\sigma^\mathbb{N}_\epsilon$ \\ \hline
    Point estimate & -0.0298 & 0.8042 & 0.0019 & 3.5199 \\ \hline
    95\% CI [lower] & -0.0414 & 0.8022 & 0.0018 & 3.3214 \\ \hline
    95\% CI [upper] & -0.0182 & 0.8061 & 0.0021 & 3.7183 \\
    \hline
    \end{tabular}}
    \caption{Regression model parameter estimates from experiment $\mathbb{E}_N$}
    \label{table:nasa_regression_results}
}
\hfill
\parbox{.6\linewidth}{
\centering
\resizebox{0.9\linewidth}{!}{
    \begin{tabular}{| l | l | l | l | l | l | l |}
    \hline
    \diagbox{\textbf{Property}}{\textbf{Coeff.}} & $\alpha^\mathbb{C}$ & $\beta^\mathbb{C}$ & $\gamma^\mathbb{C}$ & $\delta^\mathbb{C}$ & $\tau^\mathbb{C}$ & $\sigma^\mathbb{C}_\epsilon$ \\ \hline
	Point estimate & 2.4667 & 0.7679 & -0.0048 & 0.0045 & 0.0001 & 3.9678 \\ \hline
	95\% CI [lower] & 2.4399 & 0.7658 & -0.0056 & 0.0024 & -0.0001 & 3.7441 \\ \hline
	95\% CI [upper] & 2.4935 & 0.7700 & -0.0040 & 0.0066 & 0.0003 & 4.1916 \\
    \hline
	\end{tabular}}
    \caption{Regression model parameter estimates from experiment $\mathbb{E}_C$}
    \label{table:cmu_regression_results}
}
\end{table}
\subsection{Discussion}
\label{ssec:discussion}
The findings from experiment $\mathbb{E_N}$ reveal that the dashboard feedback offers a statistically significant ($p=4.16\times10^{-4}$) reduction in office plugload power consumption. The average reduction is $9.52\%$, and the $95\%$ confidence interval is $[4.32\%,14.71\%]$. This reduction is lower than the $15-40\%$ commercial office energy savings estimated in \cite{mercier2011commercial,acker2012office} or the $12\%-20\%$ reduction deemed possible by behavioral modification in \cite{yun2013sustainability}. Similarly, the data from experiment $\mathbb{E_C}$ also suggests that dashboard feedback results in a statistically significant ($p=0.03$) power reduction in the university environment. The average reduction is $21.61\%$, and the $95\%$ confidence interval is $[2.59\%,40.63\%]$. This reduction is lower than the $30\%$ reduction reported by a previous study within a university environment \cite{yun2013toward}. Our estimates may be considered conservative since they reflect power reduction only through active behavioral changes as noted in Section \ref{sssec:inactivity_threshold}. We also note that the mean reduction in the government office ($9.52\%$) is tangibly different than that of the university environment ($21.61\%$). This difference could be the result of differing power reduction potentials driven by the device lists in both environments as shown in Table \ref{table:device_list}. The office environment consisted of two desktops compared to eight at the university. Desktop computers are known to be among the largest occupant plugloads with a substantial reduction potential \cite{mercier2011commercial}. Consequently, the power reduction realized at the university environment is larger than that of the office environment. In this manner, the compressibility of energy demand, or lack thereof, determines the scope for power reduction.

The findings from experiment $\mathbb{E_C}$ suggest that the incentive, dashboard feedback, and their combination resulted in a mean reduction of $12.93$\% ($p=0.11$), $21.61\%$ ($p=0.03$), and $24.22\%$ ($p=0.02$), respectively. It is noteworthy that the incentive intervention corresponds to a larger p-value and hence less significant than the dashboard or the combined intervention. A possible explanation is to consider the order of interventions. The first experiment phase $\mathcal{P}2^\mathbb{C}$ consisted of the incentive and the later phases $\mathcal{P}3^\mathbb{C}$ and $\mathcal{P}4^\mathbb{C}$ consisted of the feedback and the combination, respectively. The growth of practical and statistical significance in the order of phases is suggestive of the effect of time on plugload power consumption behavior. This suggestion is consistent with the finding that behavioral changes require adaptation time prior to the formation of habits \cite{lally2010habits}. From the point of sustainable energy reduction, it is desirable that the interventions are instrumental in cultivating habits that persist over time \cite{frey2014persistence}. To account for the temporal effect on energy behavior, future studies could include an adaption or settle-in time during the experiment design. Further, to examine persistence and fade-out effects from one phase to another, a sufficient washout period and re-baselining need to be considered. While these considerations increase the experiment duration, they allow for a systematic investigation of the temporal effects of interventions on occupant power consumption behavior.

The ARX model coefficients for both experiments are provided in tables \ref{table:nasa_regression_results} and \ref{table:cmu_regression_results}. In both models, the coefficients $\beta^\mathbb{N}=0.8042$ and $\beta^\mathbb{C}=0.7679$ indicate significant predictive information in the lagged power differential. Given the models' accuracies, the lower effect size magnitudes of $\gamma^\mathbb{C}$, $\delta^\mathbb{N}$, $\delta^\mathbb{C}$, and $\tau^\mathbb{C}$ suggest the need to track other variables to model intervention effects better. Ideally, these variables would offer insights into participant attention, perception, and behavior pertaining to plugload energy consumption. The difference in magnitudes of the constant terms $\alpha^\mathbb{N}=-0.0298$ and $\alpha^\mathbb{C}=2.4667$ can be understood by considering the underlying AutoRegressive (AR(1)) processes in the absence of interventions. Let $\mathbb{Q} \in \{\mathbb{N}, \mathbb{C}\}$. Given that $\lvert \beta^\mathbb{Q} \rvert <1$\footnote{In the strict sense, this inequation is an assumption about the population parameter $\beta^\mathbb{Q}$ by observing that its 95\% CI estimates in tables \ref{table:nasa_regression_results} and \ref{table:cmu_regression_results} are less than 1 in magnitude. In general, the numbers associated with the regression coefficients (parameters) in this work represent their sample estimates.}, we can treat these AR(1) processes as Wide Sense Stationary (WSS) and write $\alpha^\mathbb{Q} = (1-\beta^\mathbb{Q})\mu_{\Delta Y_{\mathbb{Q}_:}}$\cite{hamilton2020time}, where $\beta^\mathbb{Q}$ and $\mu_{\Delta Y_{\mathbb{Q}_:}}$ are respectively the first order autocorrelation and mean of the hourly differential $\Delta Y_{\mathbb{Q}_:}$. Thus, the magnitudes of $\alpha^\mathbb{N}$ and $\alpha^\mathbb{C}$ differ due to the differences between participants' aggregate energy behavior in both environments.

The training and test RMS residuals of both ARX models are similar as noted in sections \ref{ssec:nasa_regression_results} and \ref{ssec:cmu_regression_results}. This similarity suggests that the models are not overfitting. Further, the test set residual plots are shown in figures \ref{FIG:nasa_residuals} and \ref{FIG:cmu_residuals}. The residual behavior does not suggest heteroscedasticity. From figures \ref{FIG:plugload_nasa_autocorr_lags} and \ref{FIG:plugload_cmu_autocorr_lags}, we note that the residual process is not autocorrelated. To the extent each of the underlying Gaussian error processes $\epsilon^\mathbb{N}$ and $\epsilon^\mathbb{C}$ are uncorrelated and homoscedastic, the respective OLS estimators can be regarded as the Best Linear Unbiased Estimator (BLUE) based on the Gauss-Markov theorem. The RMS accuracy of the office and university models on the test set are $77.39\%$ and $76.38\%$, respectively. The prediction error can be a product of factors related to modeling, estimation, or observation errors \cite{shmueli2010explain}. The observed significant reduction in plugload power consumption could be the result of behavioral changes induced either by the dashboards or cognitive factors such the Hawthorne effect where the awareness of the participants about their energy consumption being monitored changes their energy behavior \cite{kosonen2016quantifying}.

While the experiments are carried out in different buildings, the methodology to elicit the influence of visual and incentive interventions on occupant plugload usage is generalizable across buildings. We suggest the future studies to consider, (1) larger timescale and sample size with appropriate settle-in and washout periods, and (2) multimodal tracking of participant behavior to obtain further insights into occupant plugload energy consumption, including the persistence and fade-out effects of interventions. Such insights, apart from verifying the extent of generalizability, enable more accurate models comprising significant predictive information. 
\section{Conclusion}
\label{conclusion}
In this work, the problem of improving building energy efficiency through plugload management is considered. The changes in occupant plugload energy consumption due to monetary incentives and/or visual feedback were investigated by conducting experiments in office and university buildings. These experiments employed a matched pairs design to strengthen the causal connection between plugload consumption and the corresponding intervention used. During different phases of the experiments, interventions in the form of monetary incentives and/or dashboard feedback were provided. The incentives were offered in a random order and the dashboard was constructed with regard to occupant engagement and plugload consumption awareness. The experiment in the office environment was conducted at NASA Sustainability Base in the presence of dashboard feedback. The average plugload reduction was found to be $9.52\%$ ($p=4.61\times10^{-4}$) and the regression model RMS accuracy was found to be $77.39\%$. The experiment in the university environment was conducted at the CMU Silicon Valley campus in the presence of incentives and/or dashboard feedback. The average plugload reduction in the presence of incentives, dashboard feedback, and their combination was found to be $12.93$\% ($p=0.11$), $21.61\%$ ($p=0.03$), and $24.22\%$ ($p=0.02$), respectively. The regression model RMS accuracy for the university experiment was found to be $76.38\%$. Findings from both experiments indicate that feedback intervention can be effective in both office and university environments with an estimated mean reduction of $9.52\%$ and $21.61\%$, respectively. The proposed models potentially enable the integration of occupant plugload consumption control into demand response paradigms for achieving a low-carbon society. Future studies should investigate experiment designs with larger sample sizes, persistence of effects, load shifting mechanisms, sustainable interventions, and generalizability across building types and cities.
\section*{Acknowledgments}
The authors thank the NASA Ames Research Center, Carnegie Mellon University (CMU), and their Institutional Review Boards (IRBs) for supporting this research under the cooperative agreement NNX13AD49A.

{\footnotesize\bibliography{references}}

\begin{thebibliography}{10}
\expandafter\ifx\csname url\endcsname\relax
  \def\url#1{\texttt{#1}}\fi
\expandafter\ifx\csname urlprefix\endcsname\relax\def\urlprefix{URL }\fi
\expandafter\ifx\csname href\endcsname\relax
  \def\href#1#2{#2} \def\path#1{#1}\fi

\bibitem{iea2019buildingsrole}
{IEA, Buildings}, The critical role of buildings, International Energy Agency
  (2019).

\bibitem{iea2020buildingstracking}
{IEA, Paris}, Tracking buildings 2020, International Energy Agency (2020).

\bibitem{hashempour2020energy}
N.~Hashempour, et~al., Energy performance optimization of existing buildings: A
  literature review, Sustainable Cities and Society (2020).

\bibitem{longo2019review}
S.~Longo, F.~Montana, E.~R. Sanseverino, A review on optimization and
  cost-optimal methodologies in low-energy buildings design and environmental
  considerations, Sustainable cities and society 45 (2019) 87--104.

\bibitem{useia2020annual}
{US Energy Information Administration}, Annual energy outlook 2020: With
  projections to 2050 (2020).

\bibitem{nrelplugloads}
NREL, {Assessing and Reducing Plug and Process Loads in Office Buildings}, US
  Department of Energy (2010).

\bibitem{mckenney2010commercial}
K.~McKenney, et~al., {Commercial miscellaneous electric loads: Energy
  consumption characterization and savings potential in 2008 by building type},
  TIAX LLC, Lexington, MA, Tech. Rep.D (2010).

\bibitem{sehar2017integrated}
F.~Sehar, M.~Pipattanasomporn, S.~Rahman, Integrated automation for optimal
  demand management in commercial buildings considering occupant comfort,
  Sustainable cities and society 28 (2017) 16--29.

\bibitem{castleton2010green}
H.~F. Castleton, et~al., Green roofs; building energy savings and the potential
  for retrofit, Energy and buildings (2010).

\bibitem{salsbury2005survey}
T.~I. Salsbury, A survey of control technologies in the building automation
  industry, IFAC Proceedings Volumes 38~(1) (2005) 90--100.

\bibitem{wen2011personalized}
Y.-J. Wen, et~al., {Personalized dynamic design of networked lighting for
  energy-efficiency in open-plan offices}, Energy and Buildings (2011).

\bibitem{kaneda2010plug}
D.~Kaneda, B.~Jacobson, P.~Rumsey, R.~Engineers, {Plug load reduction: The next
  big hurdle for net zero energy building design}, in: ACEEE Summer Study on
  Energy Efficiency in Buildings, 2010, pp. 120--130.

\bibitem{agdas2015energy}
D.~Agdas, R.~S. Srinivasan, K.~Frost, F.~J. Masters, Energy use assessment of
  educational buildings: Toward a campus-wide sustainable energy policy,
  Sustainable Cities and Society 17 (2015) 15--21.

\bibitem{mercier2011commercial}
C.~Mercier, L.~Moorefield, {Commercial Office Plug Load Savings and Assessment:
  Executive Summary}, California Energy Commission (2011).

\bibitem{acker2012office}
B.~Acker, C.~Duarte, K.~Van Den~Wymelenberg, {Office space plug load profiles
  and energy saving interventions}, Proc. of the 2012 ACEEE Summer Study on
  Energy Efficiency in Buildings, Pacific Grove, CA (2012).

\bibitem{jain2012assessing}
R.~K. Jain, et~al., {Assessing eco-feedback interface usage and design to drive
  energy efficiency in buildings}, Energy and buildings (2012).

\bibitem{yun2013sustainability}
R.~Yun, et~al., {Sustainability in the workplace: nine intervention techniques
  for behavior change}, in: Persuasive Technology, Springer, 2013.

\bibitem{petersen2007dormitory}
J.~E. Petersen, et~al., {Dormitory residents reduce electricity consumption
  when exposed to real-time visual feedback and incentives}, International
  Journal of Sustainability in Higher Education 8~(1) (2007) 16--33.

\bibitem{GANDHI20161}
P.~Gandhi, G.~S. Brager, {Commercial office plug load energy consumption trends
  and the role of occupant behavior}, Energy and Buildings (2016).

\bibitem{bourdeau2019modeling}
M.~Bourdeau, X.~qiang Zhai, E.~Nefzaoui, X.~Guo, P.~Chatellier, Modeling and
  forecasting building energy consumption: A review of data-driven techniques,
  Sustainable Cities and Society 48 (2019) 101533.

\bibitem{nasa_sb}
NASA, {NASA Sustainability base},
  \url{http://www.nasa.gov/ames/facilities/sustainabilitybase} (2016).

\bibitem{poolla2019designing}
C.~Poolla, A.~K. Ishihara, R.~Milito, {Designing near-optimal policies for
  energy management in a stochastic environment}, Applied Energy (2019).

\bibitem{lally2010habits}
P.~Lally, C.~H. Van~Jaarsveld, H.~W. Potts, J.~Wardle, {How are habits formed:
  Modelling habit formation in the real world}, European journal of social
  psychology 40~(6) (2010) 998--1009.

\bibitem{enmetric}
{Enmetric Plug Load Management System},
  \url{https://www.enmetric.com/platform}, accessed: 2016-10-29 (2016).

\bibitem{montgomery2008design}
D.~C. Montgomery, {Design and analysis of experiments}, John Wiley \& Sons,
  2008.

\bibitem{yun2013toward}
R.~Yun, et~al., {Toward the design of a dashboard to promote environmentally
  sustainable behavior among office workers}, in: Persuasive Technology,
  Springer, 2013, pp. 246--252.

\bibitem{gulbinas2014effects}
R.~Gulbinas, J.~E. Taylor, {Effects of real-time eco-feedback and
  organizational network dynamics on energy efficient behavior in commercial
  buildings}, Energy and buildings 84 (2014) 493--500.

\bibitem{brath2004dashboard}
R.~Brath, M.~Peters, {Dashboard design: Why design is important}, DM Direct
  (2004).

\bibitem{de2002thermal}
R.~de~Dear, et~al., Thermal comfort in naturally ventilated buildings:
  revisions to ashrae standard 55, Energy buildings (2002).

\bibitem{petkov2011motivating}
P.~Petkov, F.~K{\"o}bler, M.~Foth, H.~Krcmar, {Motivating domestic energy
  conservation through comparative, community-based feedback in mobile and
  social media}, in: Proc. 5th Intl. Conf. on Communities and Technologies,
  ACM, 2011, pp. 21--30.

\bibitem{allcott2011social}
H.~Allcott, {Social norms and energy conservation}, Journal of Public Economics
  95~(9) (2011) 1082--1095.

\bibitem{ayres2012evidence}
I.~Ayres, S.~Raseman, A.~Shih, {Evidence from two large field experiments that
  peer comparison feedback can reduce residential energy usage}, Journal of
  Law, Economics, and Organization (2012) ews020.

\bibitem{pope2011round}
D.~Pope, U.~Simonsohn, Round numbers as goals: Evidence from baseball, sat
  takers, and the lab, Psychological science 22~(1) (2011) 71--79.

\bibitem{frey2014persistence}
E.~Frey, et~al., Persistence: How treatment effects persist after interventions
  stop, Policy Insights from the Behavioral and Brain Sciences (2014).

\bibitem{hamilton2020time}
J.~D. Hamilton, Time series analysis, Princeton university press, 2020.

\bibitem{shmueli2010explain}
G.~Shmueli, {To explain or to predict?}, Statistical science 25~(3) (2010)
  289--310.

\bibitem{kosonen2016quantifying}
H.~Kosonen, A.~Kim, Quantifying plug load energy use in a leed gold
  building—lessons learned in the installation phase, in: Construction
  Research Congress 2016, 2016, pp. 1234--1243.

\end{thebibliography}

\end{document}